\documentclass{ws-rv9x6}
\usepackage[square]{ws-rv-van}             

\makeindex
\begin{document}

\chapter{TASI 2006 Lectures on Leptogenesis \label{chen}}

\author[M.-C. Chen]{Mu-Chun Chen} 

\address{Theoretical Physics Department, Fermi National Accelerator Laboratory\\
Batavia, IL 60510-0500, U.S.A.\\
and\\
Department of Physics \& Astronomy, University of California\\
 Irvine, CA 92697-4575, U.S.A.\footnote{Address after January 1, 2007.}\\
muchunc@uci.edu }

\begin{abstract}
The origin of the asymmetry between matter and anti-matter of the Universe has been one of the great challenges in particle physics and cosmology. 
Leptogenesis as a mechanism for generating the cosmological baryon asymmetry of the Universe has gained significant interests ever since the advent of the evidence of non-zero neutrino masses. 
In these lectures presented at TASI 2006, I review various realizations of leptogenesis and allude to recent developments in this subject.
\end{abstract}

\body

\vspace{0.25in}
\begin{noindent}
Table of Contents
\end{noindent}

\begin{itemize}

\item[ ] Lecture I: Introduction

\begin{enumerate}

\item Evidence of Baryonic Asymmetry

\item Sakharav's Three Conditions

\item Mechanisms for Baryogenesis and Their Problems

\item Neutrino Oscillations and Sources of CP Violations

\end{enumerate}

\item[ ]Lecture II: ``Standard Scenarios''

\begin{enumerate}

\item Standard Leptogenesis (``Majorana Leptogenesis'')

\item Dirac Leptogenesis

\item Gravitino Problem

\end{enumerate}

\item[ ] Lecture III: ``Non-standard Scenarios'' and Other Issues

\begin{enumerate}

\item Resonant Leptogenesis
\item Soft Leptogenesis
\item Non-thermal Leptogenesis
\item Connection between Leptogenesis and Low Energy CP Violation
\item New Developments and Open Questions

\end{enumerate}

\end{itemize}

\section{Introduction}\label{sec:intro}

The understanding of the origin of the cosmological baryon asymmetry has been a challenge for both particle physics and cosmology. In an expanding Universe, which leads to departure from thermal equilibrium, a baryon asymmetry can be generated dynamically by charge-conjugation ($C$), charge-parity ($CP$) and baryon ($B$) number violating interactions among quarks and leptons. Possible realizations of these conditions have been studied for decades, starting with detailed investigation in the context of grand unified theories. The recent advent of the evidence of non-zero neutrino masses has led to a significant amount of work in leptogenesis. This subject is of special interests because the baryon asymmetry in this scenario is in principle entirely determined by the properties of the neutrinos. In these lectures, I  discuss some basic ingredients of leptogenesis as well as recent developments in this subject.  

These lectures are organized as follows: In Sec.~\ref{sec:intro}, I  review the basic ingredients needed for the generation of baryon asymmetry and describe various mechanisms for baryogenesis and the problems in these mechanisms. In Sec.~\ref{sec:stlpg}, I  introduce the standard leptogenesis and Dirac leptogenesis as well as the problem of gravitino over-production that exists in these standard scenarios when supersymmetry is incorporated. This is followed by Sec.~\ref{sec:alternate}, in which several alternative mechanisms that have been invented to alleviate the gravitino over-production problem are discussed. Sec.~\ref{sec:connect} focuses on the subject of connecting leptogenesis with low energy leptonic CP violating processes. 
Sec.~\ref{sec:new} concludes these lectures with discussions on the recent developments.  For exiting reviews on the subject of leptogenesis and on baryogenesis in general, see {\it e.g.} Ref.~\cite{Buchmuller:2005eh,Strumia:2006qk,Nardi:2007fs,Nir:2007zq} and \cite{Riotto:1999yt,Riotto:1998bt,Trodden:2004mj}.

\begin{figure}[t!]
\centerline{\psfig{file=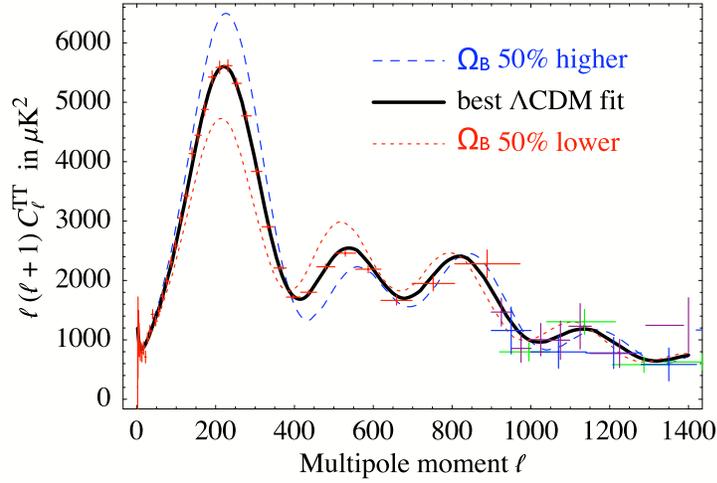,width=10.0cm}}
\caption{The power spectrum anisotropies defined in Eq.~\ref{eq:cmb1} and \ref{eq:cmb2} as a function of the multiple moment, $l$. Figure taken from Ref.~\cite{Strumia:2006qk}.}
\label{fig:wmap}
\end{figure}

\subsection{Evidence of Baryon Number Asymmetry}

\begin{figure}[b!]
\centerline{\psfig{file=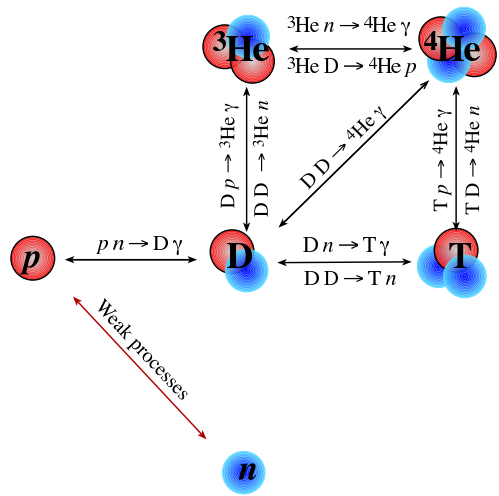,width=6.5cm}}
\caption{Main reactions that determine the primordial abundances of the light elements. Figure taken from Ref.~\cite{Strumia:2006qk}.}
\label{fig:bbn}
\end{figure}

One of the main successes of the standard early Universe cosmology is the predictions for the abundances of the light elements, D, $^{3}$He, $^{4}$He and $^{7}$Li. (For a review, see, Ref.~\cite{Copi:1994ev}. See also Scott Dodelson's lectures.) Agreement between theory and observation is obtained for a certain range of parameter, $\eta_{B}$, which is the ratio of the baryon number density, $n_{B}$, to photon density, $n_{\gamma}$,
\begin{equation}
\eta_{B}^{\mbox{\tiny BBN}} = \frac{n_{B}}{n_{\gamma}} = (2.6-6.2) \times 10^{-10} \; .
\end{equation}
The Cosmic Microwave Background (CMB) is not a perfectly isotropic radiation bath. These small temperature anisotropies are usually analyzed by 
decomposing the signal into spherical harmonics, in terms of the spherical polar angles $\theta$ and $\phi$ on the sky, as
\begin{equation}
\frac{\Delta T}{T} = \sum_{l, m} a_{lm} Y_{lm} (\theta,\phi) \; , \label{eq:cmb1}
\end{equation}
where $a_{lm}$ are the expansion coefficients. The CMB power spectrum is defined by 
\begin{equation}
C_{l} = \left< | a_{lm} |^{2} \right> \; , \label{eq:cmb2}
\end{equation}
and it is conventional to plot the quantity $l(l+1)C_{l}$ against $l$. 
The CMB measurements indicate 
that the temperature of the Universe at present is $T_{now} \sim 3^{o}K$. 
Due to the Bose-Einstein statistics, the number density of the photon, $n_{\gamma}$, scales as $T^{3}$. Together, these give a photon number density at present to be roughly $400 / \mbox{cm}^{3}$. 
It is more difficult to count the baryon number density, 
because only some fraction of the baryons form stars and other luminous objects. There are two indirect probes that point to the same baryon density. 
The measurement of CMB anisotropies probe the acoustic oscillations of the baryon/photon fluid, which happened around photon last scattering. Fig.~\ref{fig:wmap} 
illustrates how the amount of anisotropies depends on $n_{B}/n_{\gamma}$. 
The baryon number density, $n_{B} \sim 1 /\mbox{m}^{3}$, is obtained from the anisotropic in CMB, which indicates the baryon density $\Omega_{B}$ to be $0.044$. 
Another indirect probe is the Big Bang Nucleosynthesis (BBN), whose predictions depend on $n_{B}/n_{\gamma}$ through the processes shown in Fig.~\ref{fig:bbn}. 
It is measured independently from the primordial nucleosynthesis of the light elements. The value for $n_{B}/n_{\gamma}$ deduced from primordial Deuterium abundance agrees with that obtained by WMAP~\cite{Bennett:2003bz}. For ${ }^{4}$He and ${ }^{7}$Li, there are nevertheless discrepancies which may be due to the under-estimated errors. Combining WMAP measurement and the Deuterium abundance gives,  
\begin{equation}
\frac{n_{B}}{n_{\gamma}} \equiv \eta_{B} = (6.1\pm 0.3)\times 10^{-10} \; .
\end{equation}

\subsection{Sakharov's Conditions}

A matter-anti-matter asymmetry can be dynamically generated in an expanding Universe if the particle interactions and the cosmological evolution satisfy the three Sakharov's conditions~\cite{Sakharov:1967dj}: ({\it i}) baryon number violation; ({\it ii}) $C$ and $CP$ violation; ({\it iii}) departure from thermal equilibrium. 

\subsubsection{Baryon Number Violation}

As we start  from a baryon symmetric Universe ($B=0$), to evolve to a Universe where $B \ne 0$, baryon number violation is necessary. Baryon number violation occurs naturally in Grand Unified Theories (GUT), because quarks and leptons are unified in the same irreducible representations. It is thus possible to have gauge bosons and scalars mediating interactions among fermions having different baryon numbers. In the SM, on the other hand, the baryon number and the lepton number are accidental symmetries. It is thus not possible to violate these symmetries at the tree level. t'Hooft realized that~\cite{'tHooft:1976fv} the non-perturbative instanton effects may give rise to processes that violate $(B+L)$, but conserve $(B-L)$. 
Classically, $B$ and $L$ are conserved, 
\begin{equation}
B = \int d^{3} x J_{0}^{B}(x), \quad 
L = \int d^{3} x J_{0} ^{L}(x) \; ,
\end{equation}
where the currents associated with $B$ and $L$ are given by,
\begin{eqnarray}
J_{\mu}^{B} & = & \frac{1}{3} \sum_{i} \left( \overline{q}_{L_{i}} \gamma_{\mu} q_{L_{i}} - \overline{u}_{L_{i}}^{c} \gamma_{\mu} u_{L_{i}}^{c} - \overline{d}_{L_{i}}^{c} \gamma_{\mu} d_{L_{i}}^{c} \right)
\; ,
\\
J_{\mu}^{L} & = & \sum_{i} \left( \overline{\ell}_{L_{i}}\gamma_{\mu} \ell_{L_{i}} - \overline{e}_{L_{i}}^{c} \gamma_{\mu} e_{L_{i}}^{c} \right) \; .
\end{eqnarray}
Here $q_{L}$ refers to the $SU(2)_{L}$ doublet quarks, while $u_{L}$ and $d_{L}$ refer to the $SU(2)_{L}$ singlet quarks. Similarly, $\ell_{L}$ refers to the $SU(2)_{L}$ lepton doublets and $e_{L}$ refers to the $SU(2)_{L}$ charged lepton singlets. The $B$ and $L$ numbers of these fermions are summarized in Table~\ref{tbl:smf}. 
\begin{table}[t!]
\tbl{Standard model fermions and their $B$ and $L$ charges.}
{\begin{tabular}{@{}cccccccc@{}}
\toprule
\hspace{0.06in} & & $q_{L} = \left( \begin{array}{c} u \\ d \end{array} \right)_{L}$ 
 & $u_{L}^{c}$ & $d_{L}^{c}$ & $\ell_{L}=\left( \begin{array}{c} \nu \\ e \end{array} \right)_{L}$ & $e_{L}^{c}$ & \hspace{0.18in}\\
\colrule\\
& $B$ & 1/3 & -1/3 & -1/3 & 0 & 0 &\\
& $L$ & 0 & 0 & 0 & 1 & -1 & \\
\botrule
\end{tabular}}
\label{tbl:smf}
\end{table}
The subscript  $i$ is the generation index. 
Even though $B$ and $L$ are individually conserved at the tree level,  the Adler-Bell-Jackiw (ABJ) triangular anomalies~\cite{Adler:1969gk}  nevertheless do not vanish, and thus $B$ and $L$ are anomalous~\cite{Dimopoulos:1978kv} at the quantum level through the interactions with the electroweak gauge fields in the triangle diagrams (see, for example Ref.~\cite{Cheng:1985bj} for details). In other words, the divergences of the currents associated with $B$ and $L$ do not vanish at the quantum level, and they are given by
\begin{equation}
\partial_{\mu} J_{B}^{\mu} = \partial_{\mu} J_{L}^{\mu} = \frac{N_{f}}{32\pi^{2}} \left(
g^{2} W_{\mu\nu}^{p} \widetilde{W}^{p\mu\nu} - g^{\prime 2} B_{\mu\nu} \widetilde{B}^{\mu\nu}
\right) \; ,
\end{equation}
where $W_{\mu\nu}$ and $B_{\mu\nu}$ are the $SU(2)_{L}$ and $U(1)_{Y}$ field strengths, 
\begin{eqnarray}
W_{\mu\nu}^{p} & = & \partial_{\mu} W_{\nu}^{p} - \partial_{\nu} W_{\mu}^{p}\\
B_{\mu\nu} & = & \partial_{\mu} B_{\nu} - \partial_{\nu} B_{\mu} \; ,
\end{eqnarray}
respectively, with corresponding gauge coupling constants being $g$ and $g^{\prime}$, and $N_{f}$ is the number of fermion generations.
As $\partial^{\mu} (J_{\mu}^{B} - J_{\mu}^{L}) = 0$, $(B-L)$ is conserved. However,  
$(B+L)$ is violated with the divergence of the current given by,
\begin{equation}
\partial^{\mu} (J_{\mu}^{B} + J_{\mu}^{L}) = 2 N_{F} \partial_{\mu} K^{\mu} \; ,
\end{equation}
where 
\begin{eqnarray}
K^{\mu} & = &  -\frac{g^{2}}{32\pi^{2}} 2 \epsilon^{\mu\nu\rho\sigma} W_{\nu}^{p} (\partial_{\rho}W_{\sigma}^{p}  + \frac{g}{3} \epsilon^{pqr} W_{\rho}^{q}
W_{\sigma}^{r})\\
&& + \frac{g^{\prime 2} }{32\pi^{2}} \epsilon^{\mu\nu\rho\sigma} B_{\nu} B_{\rho\sigma} \; . \nonumber
\end{eqnarray} 
This violation is due to the vacum structure of non-abelian gauge theories.  
Change in $B$ and $L$ numbers are related to change in topological charges,
\begin{eqnarray}
B(t_{f}) - B(t_{i}) & = & \int_{t_{i}}^{t_{f}} dt \int d^{3}x \; \partial^{\mu} J_{\mu}^{B} \\
& = & N_{f} [N_{cs}(t_{f}) - N_{cs}(t_{i})] \nonumber \; ,
\end{eqnarray}
where the topological charge of the gauge field ({\it i.e.} the Chern-Simons number) $N_{cs}$ is given by,
\begin{equation}
 N_{cs}(t)  = \frac{g^{3}}{96 \pi^{2}} \int d^{3}x \epsilon_{ijk} \epsilon^{IJK} W^{Ii} W^{Jj} W^{Kk} \; .
\end{equation}

\begin{figure}[h!]
\centerline{\psfig{file=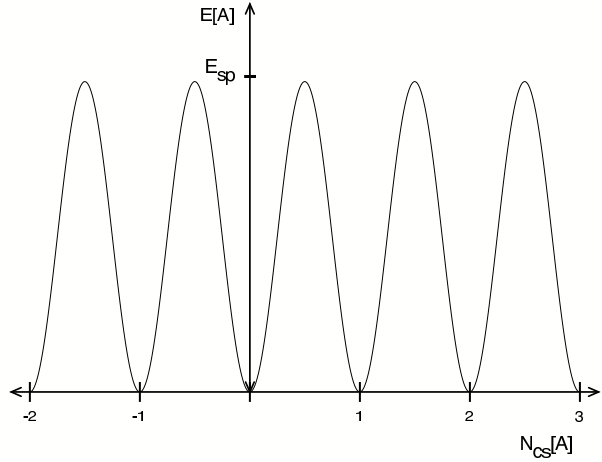,width=9.0cm}}
\caption{The energy dependence of the gauge configurations $A$ as a function of the Chern-Simons number, $N_{cs}[A]$. Sphalerons correspond to the saddle points,  {\it i.e.} maxima of the potential.}
\label{fig:potential}
\end{figure}

There are therefore infinitely many degenerate ground states with $\Delta N_{cs} = \pm 1, \; \pm 2, \; .....$, separated by a potential barrier, as depicted by Fig.~\ref{fig:potential}. 
 In semi-classical approximation, the probability of tunneling between neighboring vacua is determined by the instanton configurations.  In SM, as there are three generations of fermions, $\Delta B = \Delta L = N_{f} \Delta N_{cs} = \pm 3n$, with $n$ being an positive integer. 
 In other words, the vacuum to vacuum transition changes $\Delta B$ and $\Delta L$ by multiples of $3$ units. As a result, 
 the $SU(2)$ instantons lead to the following effective operator at the lowest order,
\begin{equation}
\mathcal{O}_{B+L} = \prod_{i=1,2,3} (q_{L_{i}}q_{L_{i}}q_{L_{i}}\ell_{L_{i}}) \; ,
\end{equation}
which gives 12 fermion interactions, such as,
\begin{equation}
\overline{u} + \overline{d} + \overline{c} \rightarrow d + 2s + 2b + t + \nu_{e} + \nu_{\mu} + \nu_{\tau} \; .
\end{equation}

At zero temperature, the transition rate is given by, $\Gamma \sim e^{-S_{int}} = e^{-4\pi / \alpha} = \mathcal{O}(10^{-165})$~\cite{'tHooft:1976fv}. The resulting transition rate is exponentially suppressed and thus it is negligible. In thermal bath, however, things can be quite different. It was pointed out by Kuzmin, Rubakov and Shaposhnikov~\cite{Kuzmin:1985mm} that, in thermal bath, the transitions between different gauge vacua  can be made not by tunneling but through thermal fluctuations over the barrier. When  temperatures are larger than the height of the barrier, the suppression due to the  Boltzmann factor disappear completely, and thus the ($B+L$) violating processes can occur at a significant rate and they can be in equilibrium in the expanding Universe. The transition rate at finite temperature in the electroweak theory is determined by the sphaleron configurations~\cite{Klinkhamer:1984di}, which are static configurations that correspond to unstable solutions  to the equations of motion. In other words, the sphaleron configurations are saddle points of the field energy of the gauge-Higgs system, as depicted in Fig.~\ref{fig:potential}. They possess Chern-Simons number equal to $1/2$ and have energy
\begin{equation}
E_{sp}(T) \simeq \frac{8\pi}{g}\left< H (T) \right> \; ,
 \end{equation}
 which is proportional to the Higgs vacuum expectation value (vev), $\left<H(T)\right>$, at finite temperature $T$. 
Below the electroweak phase transition temperature, $T < T_{EW}$, ({\it i.e.} in the Higgs phase), the transition rate 
per unit volume is~\cite{Arnold:1987mh}
\begin{equation}
 \frac{\Gamma_{B+L}}{V} = k \frac{M_{W}^{7}}{(\alpha T)^{3}} e^{-\beta E_{ph}(T)} \sim e^{\frac{-M_{W}}{\alpha kT}}  \; , 
 \end{equation}
 where $M_{W}$ is the mass of the $W$ gauge boson and $k$ is the Boltzmann constant.  The transition rate is thus still very suppressed. This result can be extrapolated to high temperature symmetric phase. It was found that, in the symmetric phase, $T \ge T_{EW}$, the transition rate 
 is~\cite{Arnold:1996dy}
 \begin{equation}
 \frac{\Gamma_{B+L}}{V} \sim \alpha^{5} \ln \alpha^{-1} T^{4} \; ,
 \end{equation} 
 where $\alpha$ is the fine-structure constant. Thus for $T > T_{EW}$, baryon number violating processes can be unsuppressed and profuse.

\subsubsection{$C$ and $CP$ Violation}

To illustrate the point that both $C$ and $CP$ violation are necessary in order to have baryogenesis, consider the case~\cite{Kolb:1988aj} in which superheavy $X$ boson have baryon number violating interactions as summarized in Table~\ref{tbl:exp1}. 
\begin{table}[h!]
\tbl{Baryon number violating decays of the superheavy $X$ boson in the toy model.}
{\begin{tabular}{@{}ccc@{}}
\toprule
process & branching fraction & $\Delta B$\\
\colrule
$X \rightarrow qq$ & $\alpha$ & 2/3\\
$X\rightarrow \overline{q} \overline{\ell}$ & $1-\alpha$ & -1/3\\
$\overline{X} \rightarrow \overline{q}\overline{q}$ & $\overline{\alpha}$ & -2/3\\
$\overline{X} \rightarrow q\ell$ & $1-\overline{\alpha}$ & 1/3\\
\botrule
\end{tabular}}
\label{tbl:exp1}
\end{table}
The baryon numbers produced by the decays of $X$ and $\overline{X}$ are,
\begin{eqnarray}
B_{X} & = & \alpha \biggl( \frac{2}{3} \biggr) + (1-\alpha)\biggl( -\frac{1}{3} \biggr)  = \alpha-\frac{1}{3} \; ,\\
B_{\overline{X}} & = & \overline{\alpha} \biggl( -\frac{2}{3} \biggr) + (1-\overline{\alpha}) \biggl( \frac{1}{3} \biggr) = - \biggl( \overline{\alpha} - \frac{1}{3} \biggr) \; ,
\end{eqnarray}
respectively. The net baryon number produced by the decays of the $X$, $\overline{X}$ pair is then,
\begin{equation}
\epsilon \equiv B_{X} + B_{\overline{X}} = (\alpha - \overline{\alpha}) \; .
\end{equation}
If $C$ or $CP$ is conserved, $\alpha = \overline{\alpha}$, it then leads to vanishing total baryon number, $\epsilon = 0$. 

\begin{figure}[b!]
\centerline{
\begin{tabular}{@{}ccc@{}}
\psfig{file=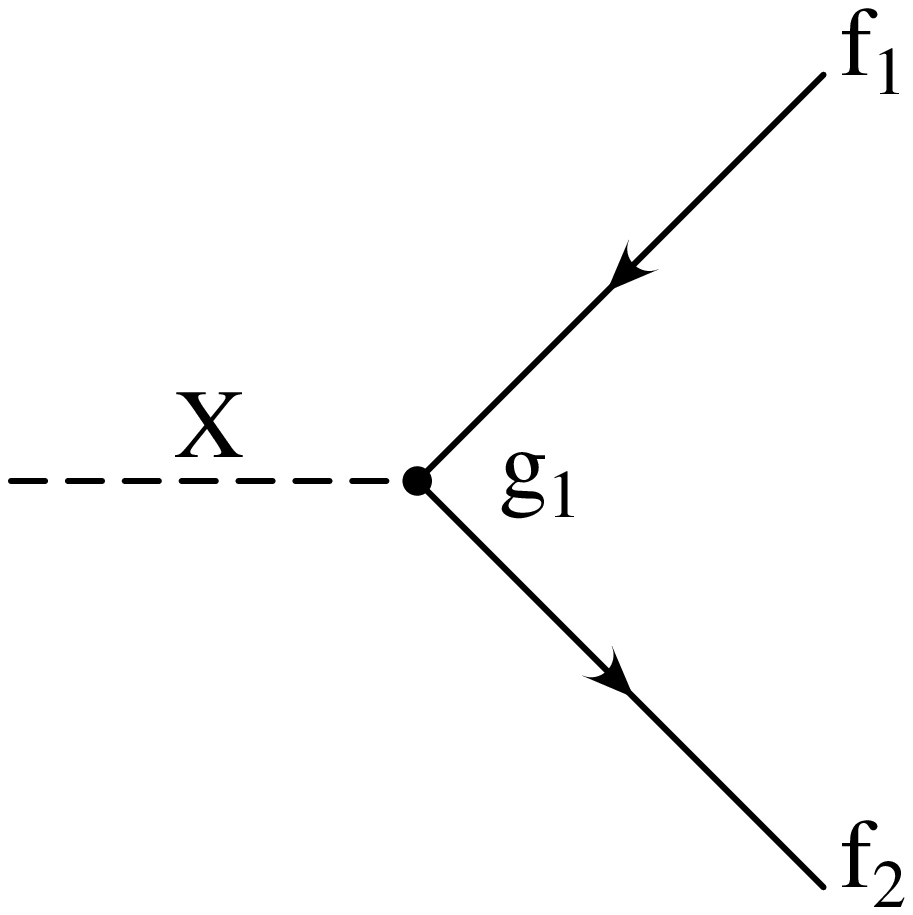,width=3.0cm} & ~~~~~~& \psfig{file=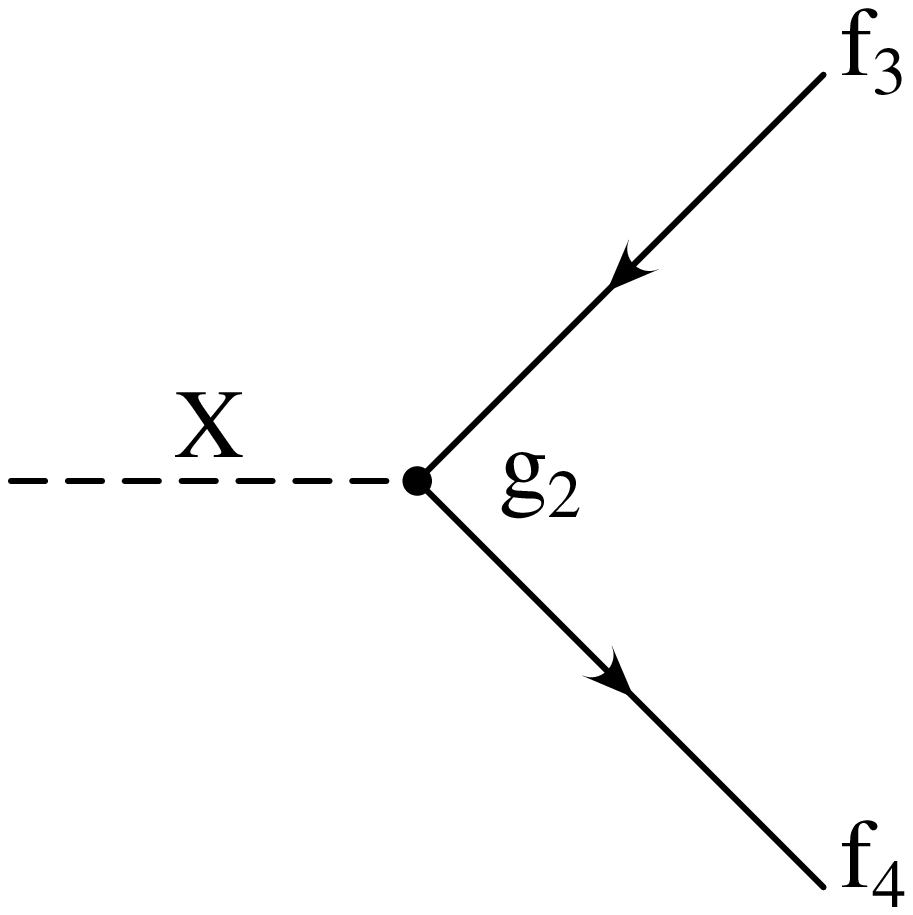,width=3.0cm}\\
(a) & & (b)\\
&&\\
\psfig{file=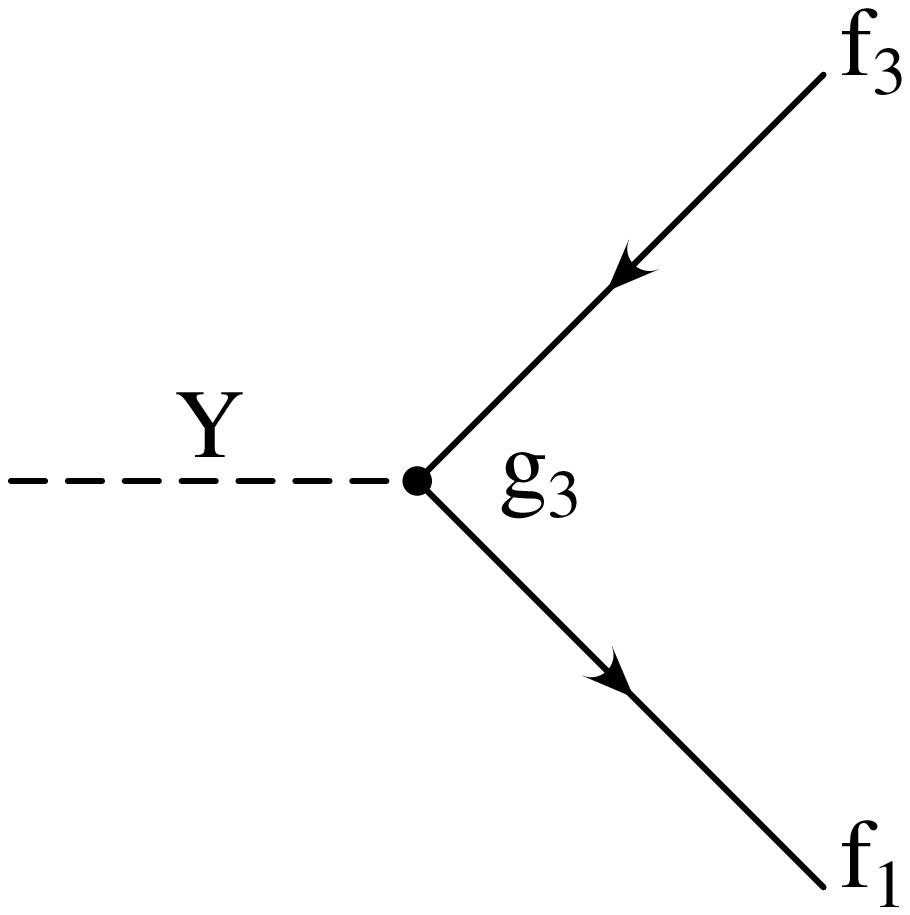,width=3.0cm} & ~~~~~ & \psfig{file=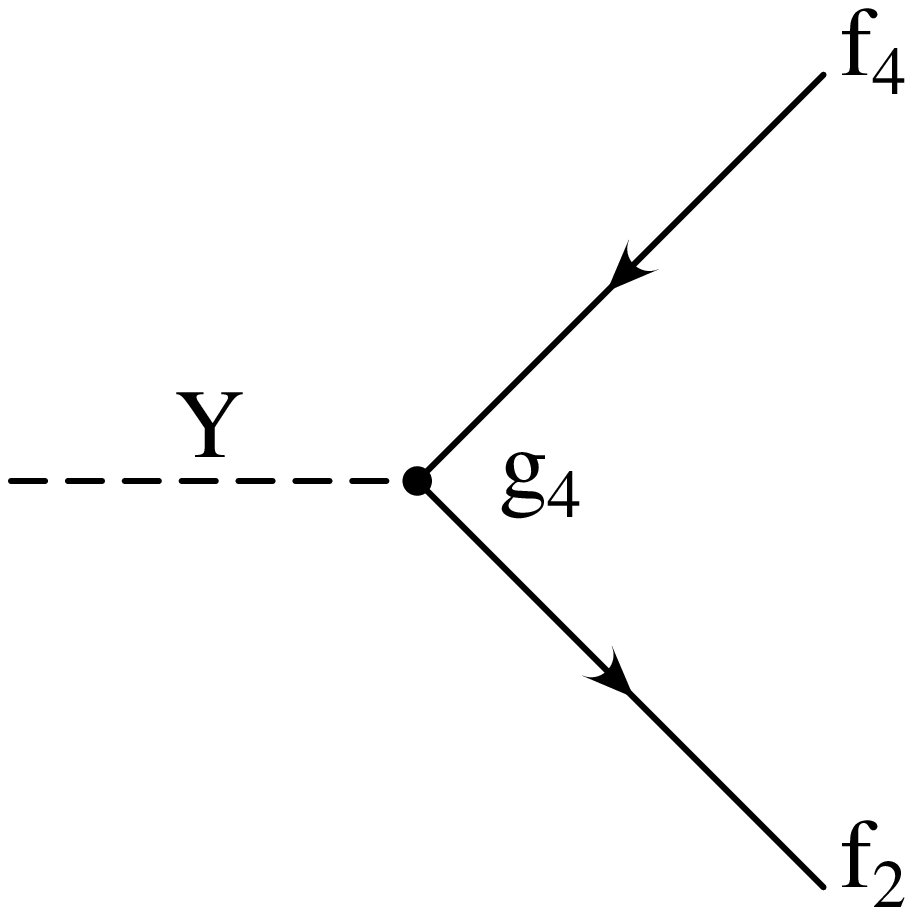,width=3.0cm}\\
(c) & & (d)
\end{tabular}}
\caption{Tree level diagrams for the decays of the heavy scalar fields,  $X$ and $Y$.}
\label{fig:tree}
\end{figure}

To be more concrete, consider a toy model~\cite{Kolb:1988aj} which consists of four fermions, $f_{1,...4}$, and two heavy scalar fields, $X$ and $Y$. The interactions among these fields are described by the following Lagrangian,
\begin{equation}
\mathcal{L} = g_{1} X f_{2}^{\dagger} f_{1} + g_{2} X f_{4}^{\dagger} f_{3} + g_{3} Y f_{1}^{\dagger} f_{3} + g_{4}Y f_{2}^{\dagger} f_{4} 
+ h.c.\; ,
\end{equation}
where $g_{1,..,4}$ are the coupling constants. The Lagrangian $\mathcal{L}$ leads to the following decay processes,
\begin{eqnarray}
X \rightarrow \overline{f}_{1} + f_{2}, \; \overline{f}_{3} + f_{4} \; , \\
Y \rightarrow \overline{f}_{3} + f_{1}, \; \overline{f}_{4} + f_{2} \; ,
\end{eqnarray}
and the tree level diagrams of these decay processes are shown in Fig.~\ref{fig:tree}. 
At the tree level, the decay rate of $X \rightarrow \overline{f_{1}} + f_{2}$ is, 
\begin{equation}
\Gamma(X\rightarrow \overline{f}_{1} + f_{2}) = |g_{1}|^{2} I_{X} \; ,
\end{equation}
where $I_{X}$ is the phase space factor. 
For the conjugate process $\overline{X} \rightarrow f_{1} + \overline{f}_{2}$, the decay rate is,
\begin{equation}
\Gamma(\overline{X} \rightarrow f_{1} + \overline{f}_{2})
= |g_{1}^{\ast}|^{2} I_{\overline{X}} \; .
\end{equation}
As the phase space factors $I_{X}$ and $I_{\overline{X}}$ are equal, no asymmetry can be generated at the tree level. 

\begin{figure}[b!]
\centerline{
\begin{tabular}{@{}ccc@{}}
\psfig{file=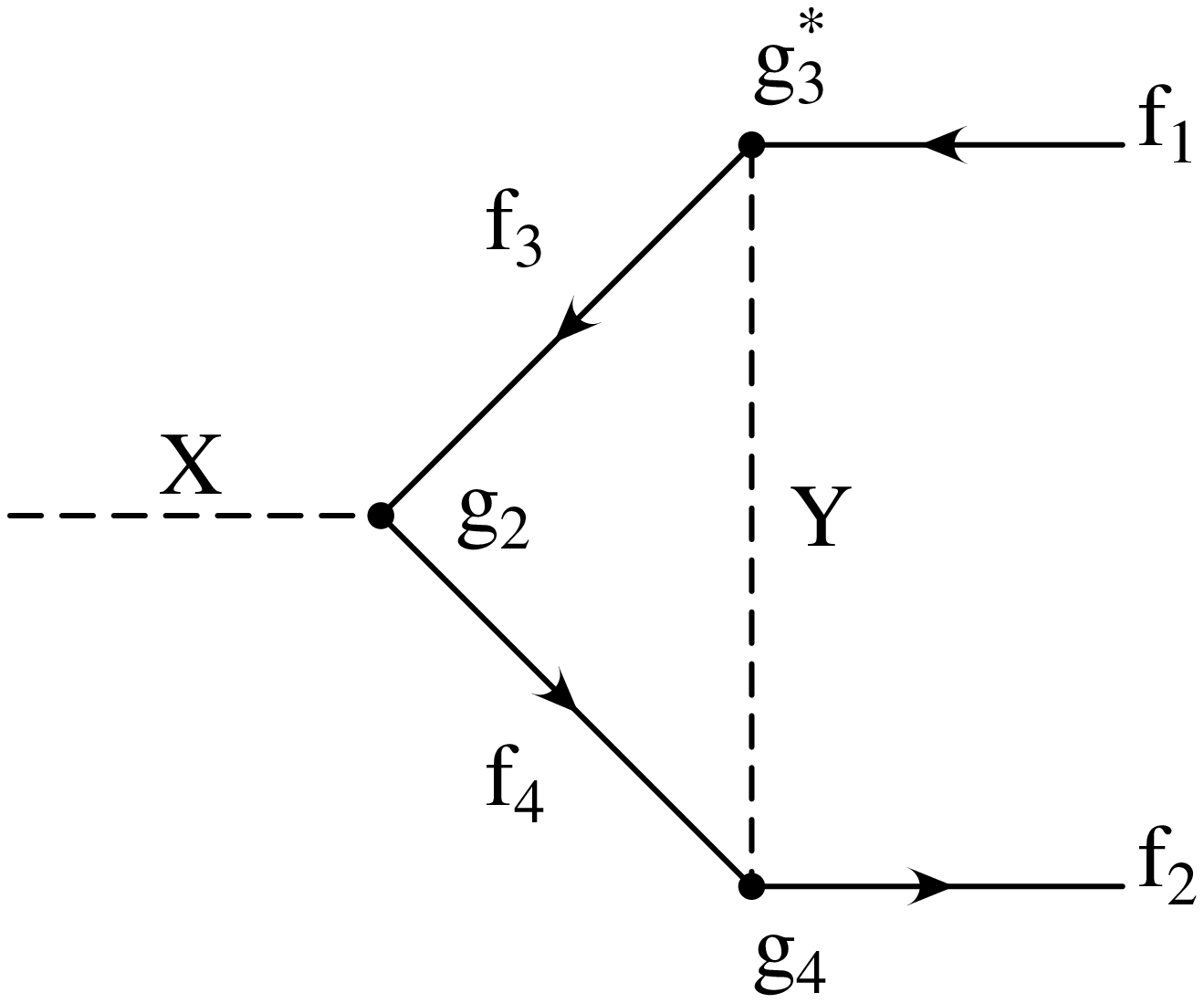,width=4.0cm} & ~~~~~~& \psfig{file=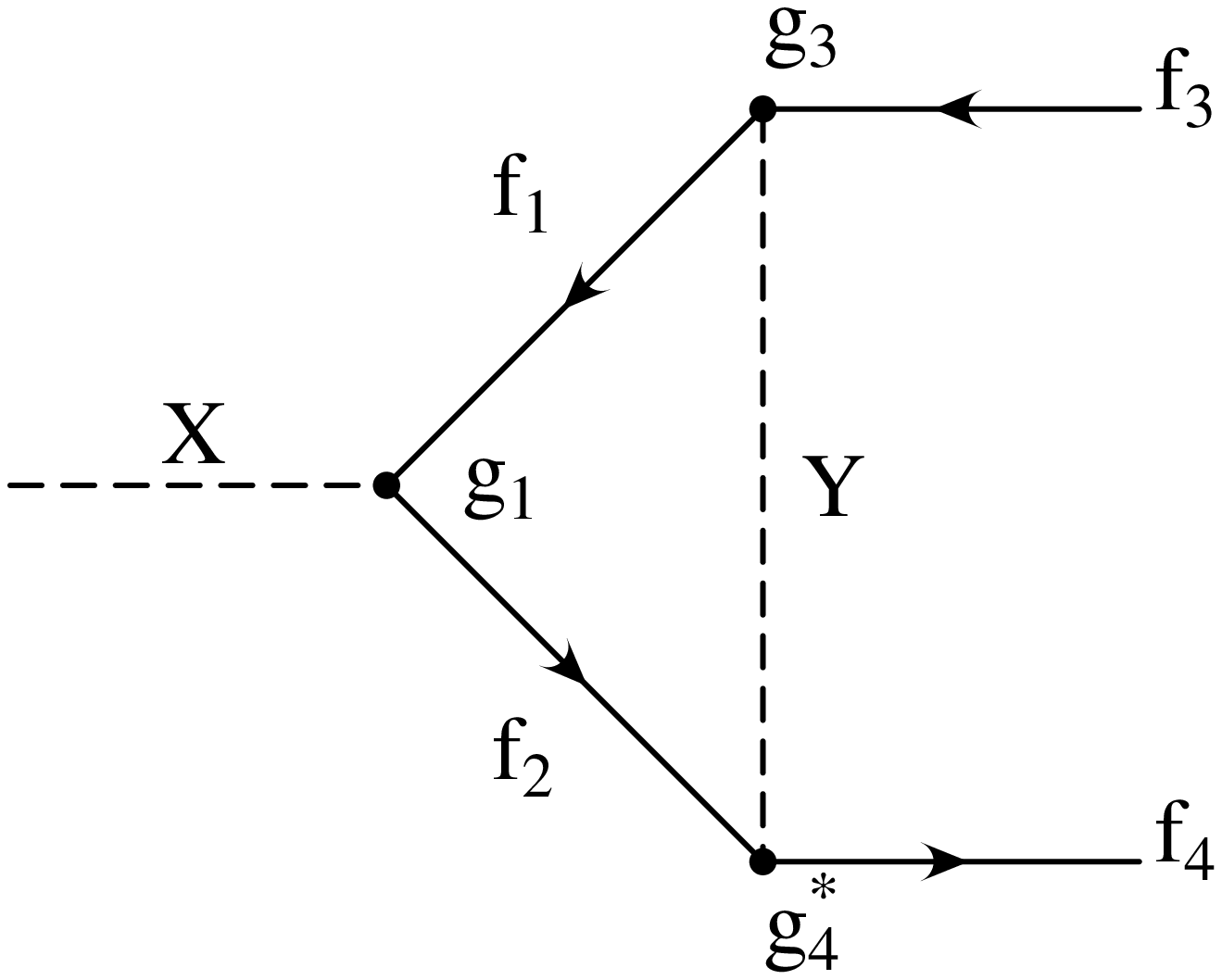,width=4.0cm}\\
(a) & & (b)\\
&&\\
\psfig{file=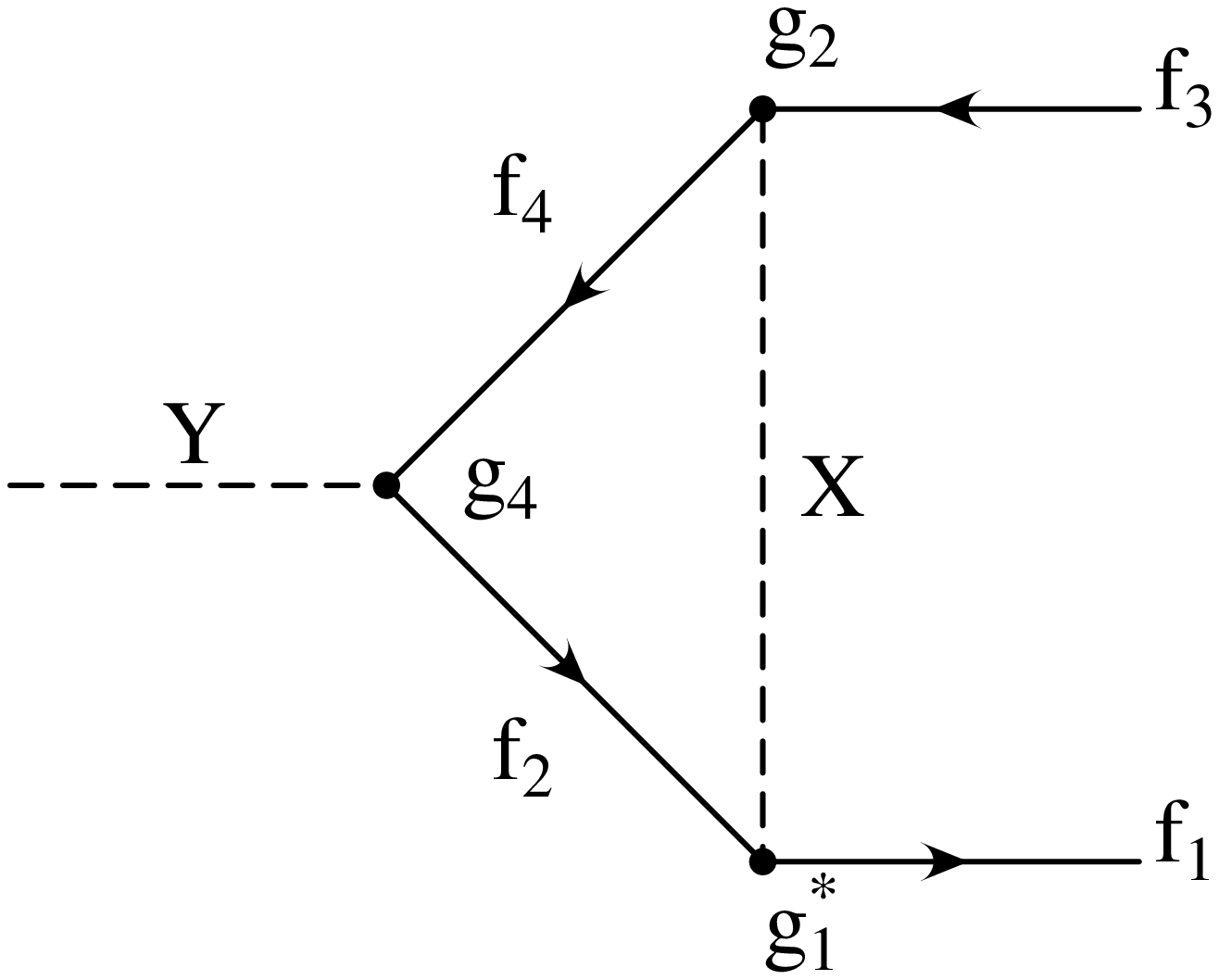,width=4.0cm} & ~~~~~ & \psfig{file=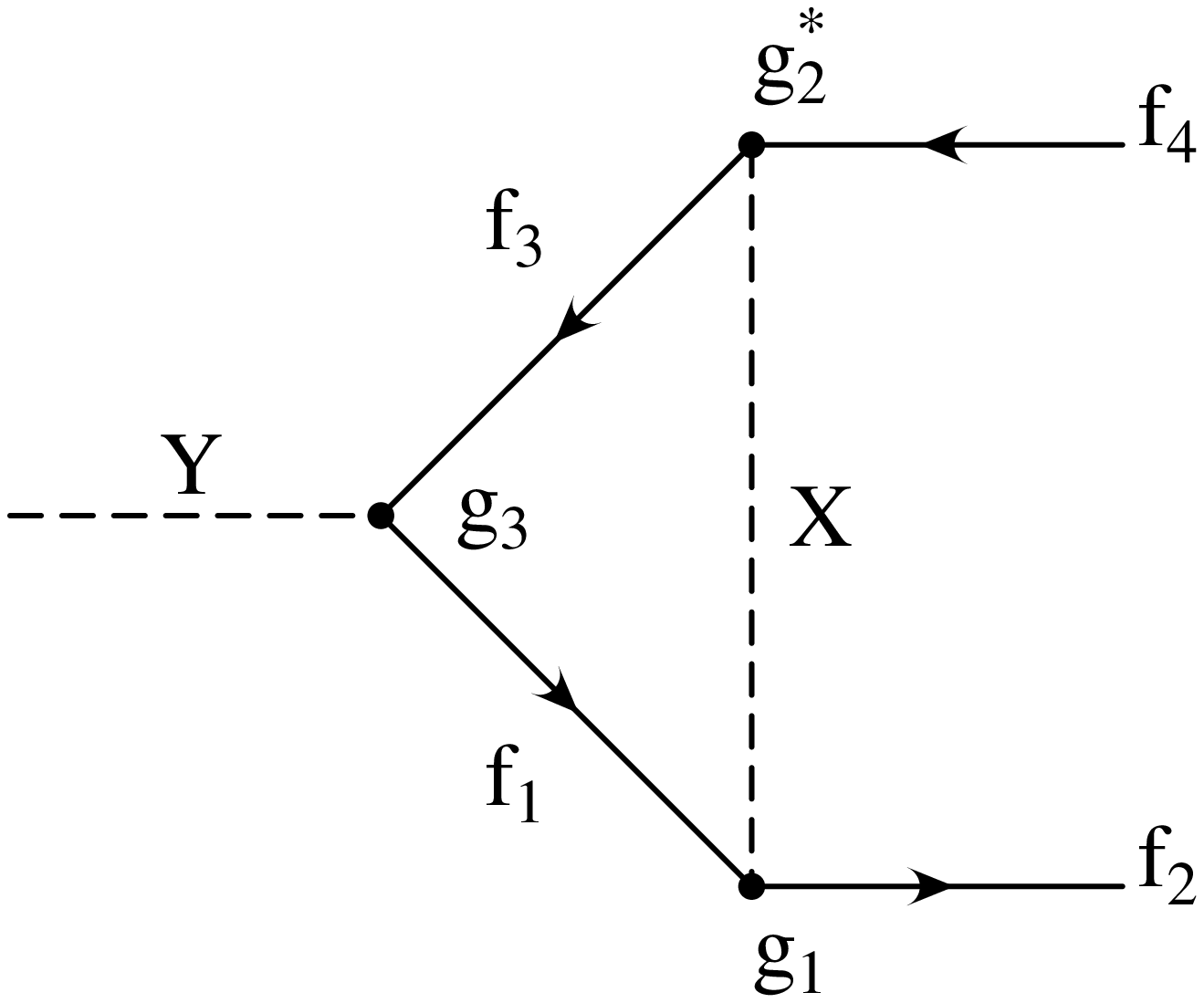,width=4.0cm}\\
(c) & & (d)
\end{tabular}}
\caption{One loop diagrams for the decays of the heavy scalar fields, $X$ and $Y$,  that contribute to the asymmetry.}
\label{fig:oneloop}
\end{figure}

At the one-loop level, there are additional diagrams, as shown in Fig.~\ref{fig:oneloop}, that have to be taken into account. 
Including these one-loop contributions, the decay rates for $X \rightarrow \overline{f}_{1} + f_{2}$ and $\overline{X} \rightarrow f_{1} + \overline{f}_{2}$ become,  
\begin{eqnarray}
\Gamma(X\rightarrow \overline{f}_{1}+f_{2}) & = & g_{1}g_{2}^{\ast}g_{3}g_{4}^{\ast} I_{XY} + c.c. \; ,
\\
\Gamma(\overline{X} \rightarrow f_{1}+\overline{f}_{2}) & = & g_{1}^{\ast} g_{2}g_{3}^{\ast}g_{4} I_{XY} + c.c. \; ,
\end{eqnarray}
where $c.c.$ stands for complex conjugation. 
Now $I_{XY}$ includes both the phase space factors as well as kinematic factors arising from integrating over the internal loop momentum due to the exchange of $J$  in $I$ decay. If fermions $f_{1,...4}$ are allowed to propagate on-shell, then the factor $I_{XY}$ is complex.
Therefore,
\begin{equation}
\Gamma(X \rightarrow \overline{f}_{1} + f_{2}) - \Gamma(\overline{X} \rightarrow f_{1} + \overline{f}_{2} )
= 4 Im(I_{XY}) Im(g_{1}^{\ast} g_{2} g_{3}^{\ast} g_{4}) \; .
\end{equation}
Similarly, for the decay mode, $X \rightarrow \overline{f}_{3} + f_{4}$, we have,
\begin{equation}
\Gamma(X \rightarrow \overline{f}_{3} + f_{4}) - \Gamma(\overline{X} \rightarrow f_{3} + \overline{f}_{4} )
= - 4 Im(I_{XY}) Im(g_{1}^{\ast} g_{2} g_{3}^{\ast} g_{4}) \; .
\end{equation}
Note that, in addition to the one-loop diagrams shown in Fig.~\ref{fig:oneloop}, there are also diagrams that involve the same boson as the decaying one. However, contributions to the asymmetry from these diagrams vanish   as the interference term in this case is proportional to $Im(g_{i}g^{\ast}_{i}g_{i}g_{i}^{\ast}) = 0$.
The total baryon number asymmetry due to $X$ decays is thus given by,
\begin{equation}
\epsilon_{X} = \frac{(B_{1} - B_{2}) \Delta \Gamma( X \rightarrow \overline{f}_{1} + f_{2})  + (B_{4} - B_{3} ) \Delta \Gamma( X \rightarrow \overline{f}_{3} + f_{4}) }{\Gamma_{X}} \; ,
\end{equation}
where
\begin{eqnarray}
\Delta \Gamma( X \rightarrow \overline{f}_{1} + f_{2}) & = &  
\Gamma(X \rightarrow \overline{f}_{1} + f_{2}) - \Gamma(\overline{X} \rightarrow f_{1} + \overline{f}_{2}) \; , \\
\Delta \Gamma( X \rightarrow \overline{f}_{3} + f_{4}) & = &
 \Gamma(X \rightarrow \overline{f}_{3} + f_{4}) - \Gamma(\overline{X} \rightarrow f_{3} + \overline{f}_{4}) \;.
\end{eqnarray}
Similar expression can be derived for the $Y$ decays. The total asymmetries due to the decays of the superheavy bosons, X and Y, are then  given, respectively, by 
\begin{eqnarray}
\epsilon_{X} & = & \frac{4}{\Gamma_{X}}Im(I_{XY}) Im(g_{1}^{\ast}g_{2}g_{3}^{\ast}g_{4}) [(B_{4}-B_{3})-(B_{2}-B_{1})] \; ,
\label{eq:epsilonx}\\
\epsilon_{Y} & = & \frac{4}{\Gamma_{Y}}Im(I_{XY}^{\prime}) Im(g_{1}^{\ast}g_{2}g_{3}^{\ast}g_{4}) [(B_{2}-B_{4})-(B_{1}-B_{3})] \; .
\label{eq:epsilony}
\end{eqnarray}

By inspecting Eq.~\ref{eq:epsilonx} and \ref{eq:epsilony}, it is clear that the following three conditions must be satisfied to have a non-zero total  asymmetry, $\epsilon = \epsilon_{X} + \epsilon_{Y}$:
\begin{itemize}
\item The presence of the two baryon number violating bosons, each of which has to have mass greater than the sum of the masses of the fermions in the internal loop; 
\item The coupling constants have to be complex. The C and CP violation then arise from  the interference between the tree level and one-loop diagrams. In general, the asymmetry generated is proportional to 
$\epsilon \sim \alpha^{n}$, with $n$ being the number of loops in the lowest order diagram that give non-zero asymmetry and $\alpha \sim g^{2}/4\pi$ ;  
\item The heavy particles $X$ and $Y$ must have non-degenerate masses. Otherwise, $\epsilon_{X} = - \epsilon_{Y}$, which leads to vanishing total asymmetry, $\epsilon$.
\end{itemize}

\subsubsection{Departure from Thermal Equilibrium}

The baryon number $B$ is odd under the $C$ and $CP$ transformations. Using this property of $B$ together with the requirement that the Hamiltonian, $H$, commutes with $CPT$, the third condition can be seen by calculating the average of $B$ 
in equilibrium at temperature $T = 1/\beta$, 
\begin{eqnarray}
<B>_{T} & = & \mbox{Tr} [e^{-\beta H} B] = \mbox{Tr} [(CPT)(CPT)^{-1}e^{-\beta H}B)] 
\\
&= & \mbox{Tr} [e^{-\beta H} (CPT)^{-1} B(CPT)] = -\mbox{Tr}[e^{-\beta H}B] \; .
\nonumber
\end{eqnarray}
In equilibrium, the average $<B>_{T} $ thus vanishes, and there is no generation of net baryon number. Different mechanisms for baryogenesis differ in the way the departure from thermal equilibrium is realized.  There are three possible ways to achieve departure from thermal equilibrium that have been utilized in baryogenesis mechanisms: 
\begin{itemize}
\item Out-of-equilibrium decay of heavy particles: GUT Baryogenesis, Leptogenesis; 
\item EW phase transition: EW Baryogenesis; 
\item Dynamics of topological defects.
\end{itemize}

\begin{figure}[b!]
\centerline{
\psfig{file=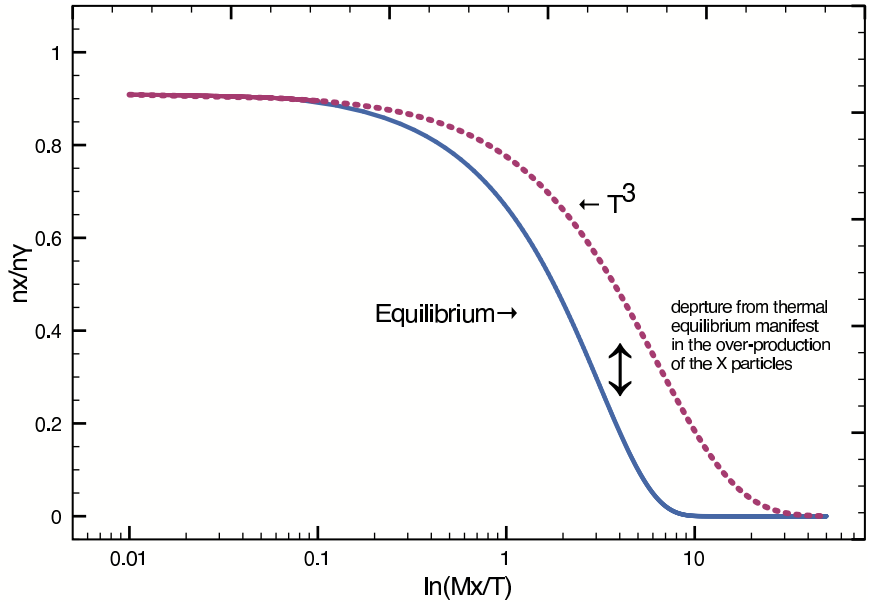,width=12.0cm}\hspace{1.2cm}}
\caption{The distribution of the $X$ particles in thermal equilibrium (blue curve) follows Eq.~\ref{eq:eq1} and \ref{eq:eq2}. When departure from the thermal equilibrium occurs, the distribution of the X particles remains the same as the thermal distribution (red dashed curve).}
\label{fig:equil}
\end{figure}

In leptogenesis, the departure from thermal equilibrium is achieved through the out-of-equilibrium decays of heavy particles in an expanding Universe. 
If the decay rate $\Gamma_{X}$ of some superheavy particles $X$ with mass $M_{X}$ at the time when they become non-relativistic ({\it i.e.} $T \sim M_{X}$) is much smaller than the expansion rate of the Universe, the $X$ particles cannot decay on the time scale of the expansion. The $X$ particles will then remain their initial thermal abundance, $n_{X} = n_{\overline{X}} \sim n_{\gamma} \sim T^{3}$, for $T \lesssim M_{X}$. In other words, at some temperature $T > M_{X}$, the superheavy particles $X$ are so weakly interacting that they cannot catch up with the expansion of the Universe. Hence they decouple from the thermal bath while still being relativistic. At the time of the decoupling, $n_{X} \sim n_{\overline{X}} \sim T^{3}$. Therefore, they populate the Universe at $T \simeq M_{X}$ with abundance much larger than their abundance in equilibrium.  Recall that in equilibrium, 
\begin{eqnarray}
n_{X} & =  n_{\overline{X}} & \simeq n_{\gamma} \quad \mbox{for} \quad  T \gtrsim M_{X}\; ,
\label{eq:eq1}\\
n_{X} & =  n_{\overline{X}} & \simeq (M_{X} T)^{3/2} e^{-M_{X}/T} \ll n_{\gamma} \quad \mbox{for} \quad T \lesssim M_{X} \; .
\label{eq:eq2}
\end{eqnarray}
This over-abundance at temperature below $M_{X}$, as shown in Fig.~\ref{fig:equil},  is the departure from thermal equilibrium needed to produce a final non-vanishing baryon asymmetry, when the heavy states, $X$, undergo $B$ and $CP$ violating decays. 
The scale of rates of these decay processes involving $X$ and $\overline{X}$ relative to the expansion rate of the Universe is determined by $M_{X}$,
\begin{equation}
\frac{\Gamma}{H} \propto \frac{1}{M_{X}} \; .
\end{equation}
The out-of-equilibrium condition, $\Gamma < H$, thus requires very heavy states: for gauge bosons, $M_{X} \gtrsim (10^{15-16})$ GeV; for scalars, $M_{X} \gtrsim (10^{10-16})$ GeV, assuming these heavy particles decay through renormalizable operators. 
Precise computation of the abundance is carried out by solving the Boltzmann equations (more details in Sec.~\ref{sec:boltz}). 

\subsection{Relating Baryon and Lepton Asymmetries}\label{sec:convert}

One more ingredient that is needed for leptogenesis is to relate lepton number asymmetry to the baryon number asymmetry, at the high temperature, symmetric phase of the SM~\cite{Buchmuller:2005eh}. In a weakly coupled plasma with temperature $T$ and volume $V$, a chemical potential $\mu_{i}$ can be assigned to each of the quark, lepton and Higgs fields, $i$. There are therefore  $5N_{f}+1$ chemical potentials in the SM with one Higgs doublet and  $N_{f}$ generations of fermions. The corresponding partition function is given by,
\begin{equation}
Z(\mu, T, V) = \mbox{Tr} [e^{-\beta (H - \sum_{i} \mu_{i} Q_{i})}] \;
\end{equation}
where $\beta = 1/T$, $H$ is the Hamiltonian and $Q_{i}$ is the charge operator for the corresponding field. The asymmetry in particle and antiparticle number densities is given by the derivative of the thermal-dynamical potential, $\Omega(\mu,T)$, as
\begin{equation}
n_{i} - \overline{n}_{i} = - \frac{\partial \Omega (\mu, T)}{\partial \mu_{i}}  \; ,
\end{equation}
where $\Omega(\mu,T)$ is defined as,
\begin{equation}
\Omega (\mu,T) = -\frac{T}{V} \ln Z(\mu,T,V) \; .
\end{equation}
For a non-interacting gas of massless particles, assuming $\beta \mu_{i} \ll 1$,
\begin{equation}\label{eq:chem1}
n_{i}-\overline{n}_{i} = \frac{1}{6} gT^{3} \bigg\{
\begin{array}{ll}
\beta \mu_{i} + \mathcal{O}((\beta\mu_{i})^{3}), \qquad &  \mbox{fermions}\\
2\beta \mu_{i} + \mathcal{O}( (\beta \mu_{i})^{3} ), \qquad & \mbox{bosons} \; .
\end{array}
\end{equation}
In the high temperature plasma, quarks, leptons and Higgs interact via the guage and Yukawa couplings. In addition, there are non-perturbative sphaleron processes. All these processes give rise to constraints among various chemical potentials in thermal equilibrium.
These include~\cite{Buchmuller:2005eh}:
\begin{enumerate}
\item The effective 12-fermion interactions $\mathcal{O}_{B+L}$ induced by the sphalerons give rise to the following relation, 
\begin{equation}
\sum_{i} (3\mu_{q_{i}} + \mu_{\ell_{i}}) = 0 \; .
\end{equation}
\item The SU(3) QCD instanton processes lead to interactions between LH and RH quarks. These interactions are described by the operator,  $\prod_{i} (q_{L_{i}}q_{L_{i}}u_{R_{i}}^{c} d_{R_{i}}^{c})$. When in equilibrium, they lead to,
\begin{equation}
\sum_{i} (2 \mu_{q_{i}} - \mu_{u_{i}} - \mu_{d_{i}}) = 0 \; .
\end{equation}
\item Total hypercharge of  the plasma has to vanish at all temperatures. This gives, 
\begin{equation}
\sum_{i} (\mu_{q_{i}} + 2 \mu_{u_{i}} - \mu_{d_{i}} - \mu_{\ell_{i}} - \mu_{e_{i}} + \frac{2}{N_{f}} \mu_{H}) = 0 \; .
\end{equation}
\item The Yukawa interactions yield the following relations among chemical potential of the LH and RH fermions,
\begin{eqnarray}
\mu_{q_{i}} - \mu_{H} - \mu_{d_{j}} & = & 0 \; , \\
\mu_{q_{i}} + \mu_{H} - \mu_{u_{j}} & = & 0 \; , \\
\mu_{\ell_{i}} - \mu_{H} - \mu_{e_{j}} & = & 0 \; .
\end{eqnarray}
\end{enumerate}
From Eq.~(\ref{eq:chem1}), the baryon number density $n_{B} = \frac{1}{6} g B T^{2}$ and lepton number density $n_{L} = \frac{1}{6} g L_{i} T^{2}$, where $L_{i}$ is the individual lepton flavor number  with $i = (e, \mu,  \tau)$, can be expanded in terms of the chemical potentials. Hence 
\begin{eqnarray}
B & = & \sum_{i} (2\mu_{q_{i}} + \mu_{u_{i}} + \mu_{d_{i}})\\
L & = & \sum_{i} L_{i}, \quad L_{i}  =  2\mu_{\ell_{i}} + \mu_{e_{i}} \; .
\end{eqnarray} 
Consider the case where all Yukawa interactions are in equilibrium. The asymmetry $(L_{i} - B/N_{f})$ is then preserved. If we further assume equilibrium among different generations, $\mu_{\ell_{i}} \equiv \mu_{\ell}$ and $\mu_{q_{i}} \equiv \mu_{q}$, together with the sphaleron and hypercharge constraints, all the chemical potentials can then be expressed in terms of $\mu_{\ell}$,
\begin{eqnarray}
\mu_{e} = \frac{2N_{f}+3}{6N_{f}+3}\mu_{\ell}, \quad \mu_{d} = -\frac{6N_{f}+1}{6N_{f}+3} \mu_{\ell}, 
\quad \mu_{u} = \frac{2N_{f}-1}{6N_{f}+3} \mu_{\ell}\\
\mu_{q} = -\frac{1}{3} \mu_{\ell}, \quad \mu_{H} = \frac{4N_{f}}{6N_{f}+3}\mu_{\ell}  \; . \nonumber
\end{eqnarray}
The corresponding $B$ and $L$ asymmetries are
\begin{eqnarray}
B & = & -\frac{4}{3} N_{f} \mu_{\ell} \; , \\
L & = & \frac{14N_{f}^{2}+ 9N_{f}}{6N_{f}+3} \mu_{\ell} \; .
\end{eqnarray}
Thus $B$, $L$ and $B-L$ are related by:
\begin{equation}
B = c_{s} (B-L), \quad L=(c_{s}-1)(B-L) \; , 
\end{equation}
where 
\begin{equation}\label{eq:convert}
c_{s} = \frac{8N_{f} + 4}{22N_{f}+13} \; .
\end{equation} 
For models with $N_{H}$ Higgses, the parameter $c_{s}$ is given by,
\begin{equation}
c_{s} = \frac{8N_{f} + 4N_{H}}{22N_{f}+13N_{H}} \; .
\end{equation} 

For $T = 100 \; \mbox{GeV} \sim 10^{12} \; \mbox{GeV}$, which is of interest of baryogenesis, gauge interactions are in equilibrium. Nervertheless, the Yukawa interactions are in equilibrium only in a more restricted temperature range. 
But these effects are generally small, and thus will be neglected in these lectures. These effects have been investigated recently; they will be discussed 
in Sec.~\ref{sec:new}.  

\subsection{Mechanisms for Baryogenesis and Their Problems}\label{sec:bg}

There have been many mechanisms for baryogenesis proposed. Each has attractive and problematic aspects, which we discuss below. 

\subsubsection{GUT Baryongenesis}

The GUT baryogenesis was the first implementation of Sakharov's $B$-number generation idea. The $B$-number violation is an unavoidable consequence in grand unified models, as quarks and leptons are unified in the same representation of a single group. Furthermore, sufficient amount of $CP$ violation can be incorporated naturally in GUT models, as there exist many possible complex phases, in addition to those that are present in the SM. The relevant time scales of the decays of heavy gauge bosons or scalars are slow, compared to the expansion rate of the Universe at early epoch of the cosmic evolution. The decays of these heavy particles are thus inherently out-of-equilibrium. 

Even though GUT models naturally encompass all three Sakharov's conditions, there are also challenges these models face. First of all, to generate sufficient baryon number asymmetry requires high reheating temperature. This in turn leads to dangerous production of relic particles, such as gravitinos 
(see Sec.~\ref{sec:gravitino}). As the relevant physics scale $M_{GUT} \sim 10^{16}$ GeV is far above the electroweak scale,  it is also very hard to test GUT models experimentally using colliders. The electroweak theory ensures that there are copious $B$-violating processes between the GUT scale and the electroweak scale. These sphelaron processes violate $B + L$, but conserve $B - L$. Therefore, unless a GUT mechanism generates an excess of $B-L$, any baryon asymmetry produced will be equilibrated to zero by the sphaleron effects. As $U(1)_{B-L}$ is a gauged subgroup of $SO(10)$, GUT models based on $SO(10)$ are especially attractive for baryogenesis.

\subsubsection{EW Baryogenesis}

In electroweak baryogenesis, the departure from thermal equilibrium is provided by strong first order phase transition. The nice feature of this mechanism is that it can be probed in collider experiments. On the other hand, the allowed parameter space is very small. It requires more $CP$ violation than what is provided in the SM.  Even though there are additional sources of $CP$ violation in MSSM, the requirement of strong first order phase transition translates into a stringent bound on the Higgs mass, $m_{H} \lesssim 120$ GeV. To obtain a Higgs mass of this order, 
the stop mass needs to be smaller than, or of the order of, the top quark mass, which implies fine-tuning in the model parameters.

\subsubsection{Affleck-Dine Baryogensis}

The Affleck-Dine baryogenesis~\cite{Affleck:1984fy} involves cosmological evolution of scalar fields which carry $B$ charges. It is 
most naturally implemented in SUSY theories. Nevertheless, this mechanism 
faces the same challenges as in GUT baryogenesis and in EW baryogenesis.

\subsection{Sources of CP Violation}\label{sec:counting}

In the SM, $C$ is maximally broken, since only LH electron couples to the $SU(2)_{L}$ gauge fields. Furthermore, $CP$ is not an exact symmetry in weak interaction, as observed in the Kaon and B-meson systems. The charged current in the weak interaction basis is given by,
\begin{equation}
\mathcal{L}_{W} = \frac{g}{\sqrt{2}} \overline{U}_{L} \gamma^{\mu} D_{L} W_{\mu} + h.c. \; ,
\end{equation}
where $U_{L} = (u, c, t)_{L}$ and $D_{L} = (d,s,b)_{L}$. Quark mass matrices can be diagonalized by bi-unitary transformations,
\begin{eqnarray}
\mbox{diag}(m_{u}, m_{c}, m_{t}) & = & V_{L}^{u} M^{u} V_{R}^{u} \; ,
\\
\mbox{diag}(m_{d}, m_{s}, m_{d}) & = & V_{L}^{d} M^{d} V_{R}^{d} \; . 
\end{eqnarray}
Thus the charged current interaction in the mass eigenstates reads,
\begin{equation}
\mathcal{L}_{W} = \frac{g}{\sqrt{2}} \overline{U}_{L}^{\prime} U_{\scriptscriptstyle CKM} \gamma^{\mu} D_{L}^{\prime} W_{\mu} + h.c. \; ,
\end{equation}
where $U_{L}^{\prime} \equiv V_{L}^{u} U_{L}$ and $D_{L}^{\prime} \equiv V_{L}^{d} D_{L}$ are the mass eigenstates, and $U_{\scriptscriptstyle CKM} \equiv V_{L}^{u} (V_{L}^{d})^{\dagger}$ is the CKM matrix. For three families of fermions, 
the unitary matrix $K$ can be parameterized by three angles and six phases. Out of these six phases, five of them can be reabsorbed by redefining the wave functions of the quarks. There is hence only one physical phase in the CKM matrix. This is the only source of CP violation in the SM. It turns out that this particular source is not strong enough to accommodate the observed matter-antimatter asymmetry. The relevant effects can be parameterized by~\cite{Shaposhnikov:1986jp},
\begin{equation}
B \simeq \frac{\alpha_{w}^{4} T^{3}}{s} \delta_{CP} \simeq 10^{-8} \delta_{CP} \; ,
\end{equation}
where $\delta_{CP}$ is the suppression factor due to CP violation in the SM. Since CP violation vanishes when any two of the quarks with equal charge have degenerate masses, a naive estimate gives the effects of CP violation of the size
\begin{eqnarray}
A_{CP}  & = & (m_{t}^{2} - m_{c}^{2}) (m_{c}^{2} - m_{u}^{2}) (m_{u}^{2} - m_{t}^{2}) \\
&& \quad \cdot 
(m_{b}^{2} - m_{s}^{2}) (m_{s}^{2} - m_{d}^{2}) (m_{d}^{2} - m_{b}^{2}) \cdot J\; .
\nonumber
\end{eqnarray}
Here the proportionality constant $J$ is the usual Jarlskog invariant, which is a parameterization independent measure of CP violation  in the quark sector. 
Together with the fact that $A_{CP}$ is of mass (thus temperature) dimension $12$,  this leads to the following value for $\delta_{CP}$, which is a dimensionless quantity,
\begin{equation}
\delta_{CP} \simeq \frac{A_{CP}}{T_{C}^{12}} \simeq 10^{-20} \; ,
\end{equation}
and $T_{C}$ is the temperature of the electroweak phase transition. 
The baryon number asymmetry due to the phase in the CKM matrix is therefore of the order of $B \sim 10^{-28}$, which is too small to account for the observed $B \sim 10^{-10}$. 

In MSSM, there are new sources of CP violation due to the presence of the soft SUSY breaking sector.  The superpotential of the MSSM is given by, 
\begin{equation}
W = \mu \hat{H}_{1} \hat{H}_{2} + h^{u} \hat{H}_{2} \hat{Q} \hat{u}^{c} + h^{d} \hat{H}_{1} \hat{Q} \hat{d}^{c} 
+ h^{e} \hat{H}_{1} \hat{L} \hat{e}^{c} \; .
\end{equation}  
The soft SUSY breaking sector has the following parameters:
\begin{itemize}
\item tri-linear couplings: $\Gamma^{u} H_{2} \widetilde{Q} \widetilde{c}^{c} + \Gamma^{d} H_{1} \widetilde{Q} \widetilde{d}^{c} 
+ \Gamma^{e} H_{1} \widetilde{L} \widetilde{e}^{c} + h.c.$, where $\Gamma^{(u,d,e)} \equiv A^{(u,d,e)} \cdot h^{(u,d,e)}$; 
\item bi-linear coupling in the Higgs sector: $\mu B H_{1} H_{2}$; 
\item gaugino masses: $M_{i}$ for $i=1,2,3$ (one for each gauge group); 
\item soft scalar masses: $\widetilde{m}_{f}$. 
\end{itemize} 
In the constrained MSSM (CMSSM) model with mSUGRA boundary conditions at the GUT scale, a universal value is assumed for the tri-linear coupling constants, $A^{(u,d,e)} = A$. Similarly, the gaugino masses and scalar masses are universal, $M_{i} = M$, and $\widetilde{m}_{f} = \widetilde{m}$. 
Two phases may be removed by redefining the phase of $\hat{H}_{2}$ such that the phase of $\mu$ is opposite to the phase of $B$. As a result, the product 
$\mu B$ is real. Furthermore, the phase of $M$ can be removed by R-symmetry transformation. This then modifies the tri-linear couplings by an additional factor of 
$e^{-\phi_{M}}$, while other coupling constants are invariant under the R-symmetry transformation. There are thus two physical phases remain, 
\begin{equation}
\phi_{A} = \mbox{Arg}(AM), \quad \phi_{\mu} = -\mbox{Arg}(B) \; .
\end{equation}
These phases are relevant in soft leptogenesis, which is discussed in Sec.~\ref{sec:softlept}.

If the neutrinos are massive, the leptonic charged current interaction in the mass eigenstates of the leptons is given by,
\begin{eqnarray}
\mathcal{L}_{W} 
& = & \frac{g}{\sqrt{2}} \overline{\nu}_{L}^{\prime} U_{\scriptscriptstyle MNS}^{\dagger} \gamma^{\mu} \ell_{L}^{\prime} W_{\mu} + h.c. \; ,
\end{eqnarray}
where $U_{\scriptscriptstyle MNS} =  (V_{L}^{\nu})^{\dagger}V_{L}^{e}$. (For a review on physics of the massive neutrinos, see, {\it e.g.} 
Ref.~\cite{Chen:2003zv} and \cite{deGouvea:2004gd}. See also Rabi Mohapatra's lectures.) The matrices $V_{L}^{\nu}$ and $V_{L}^{e}$ diagonalize the effective neutrino mass matrix and the charged lepton mass matrix, respectively. 
If neutrinos are Majorana particles, which is the case if small neutrino mass is explained by the seesaw mechanism~\cite{seesaw}, 
the Majorana condition then forbids the phase redefinition of $\nu_{R}$. Unlike in the CKM matrix, in this case only three of the six complex phases can be absorbed, and there are thus two additional physical phases in the lepton sector if neutrinos are Majorana fermions. And due to this reason, CP violation can occur in the lepton sector with only two families. (Recall that in the quark sector, CP violation can occur only when the number of famalies is at least three). The MNS matrix can be parameterized as a CKM-like matrix and a diagonal phase matrix,  
\begin{eqnarray}
U_{MNS} & = & \left(
\begin{array}{ccc}
c_{12} c_{13} & s_{12} c_{13} & s_{13} e^{-i\delta}
\\
-s_{12} c_{23} - c_{12} s_{23} s_{13} e^{i\delta}
& 
c_{12} c_{23} - s_{12} s_{23} s_{13} e^{i\delta}
&
s_{23} c_{13} 
\\
s_{12} s_{23} - c_{12} c_{23} s_{13} e^{i\delta}
&
-c_{12} s_{23} - s_{12} c_{23} s_{13} e^{i\delta}
&
c_{23} c_{13}
\end{array}\right) \nonumber
\\
& & \qquad \cdot
\left(\begin{array}{ccc}
1 & &
\\ 
& e^{i\alpha_{21}/2} &
\\ 
& & e^{i\alpha_{31}/2}
\end{array}\right) \; .
\end{eqnarray}
The Dirac phase $\delta$ affects neutrino oscillation (see Boris Kayser's lectures),  
\begin{eqnarray}
P(\nu_{\alpha}\rightarrow \nu_{\beta})
& = & \delta_{\alpha\beta} - 4 \sum_{i>j} 
Re(U_{\alpha i} U_{\beta j} U_{\alpha j}^{\ast} U_{\beta i}^{\ast}) 
\sin^{2}\biggl(\Delta m_{ij}^{2} \frac{L}{4E}\biggr) 
\\ &&
+ 2 \sum_{i>j} J_{\mbox{\tiny CP}}^{\mbox{\tiny lep}} \sin^{2} \biggl( \Delta m_{ij}^{2} \frac{L}{4E} \biggr) 
\nonumber
\end{eqnarray}
where the parameterization invariant CP violation measure, the leptonic Jarlskog invariant $J_{\mbox{\tiny CP}}^{\mbox{\tiny lep}}$, is given by,
\begin{equation}
J_{\mbox{\tiny CP}}^{\mbox{\tiny lep}} = - \frac{Im(H_{12} H_{23} H_{31})}{\Delta m_{21}^{2} \Delta m_{32}^{2} \Delta m_{31}^{2}}, 
\quad
H \equiv (M_{\nu}^{eff}) (M_{\nu}^{eff})^{\dagger} \; .
\end{equation}
The two Majorana phases, $\alpha_{21}$ and $\alpha_{31}$, affect neutrino double decay (see Petr Vogel's lectures). Their dependence in the neutrinoless double beta decay matrix element is,
\begin{eqnarray}
\left| \left< m_{ee} \right>\right|^{2} & = & 
m_{1}^{2} \left| U_{e1}\right|^{4} + m_{2}^{2} \left| U_{e2} \right|^{4} + m_{3}^{2} \left| U_{e3} \right|^{4} 
\\
&& + 2 m_{1} m_{2} \left| U_{e1} \right|^{2} \left| U_{e2} \right|^{2} \cos\alpha_{21}
\nonumber\\
&&+ 2m_{1}m_{3} \left| U_{e1} \right|^{2} \left| U_{e3} \right|^{2} \cos\alpha_{31}
\nonumber\\
&&+ 2 m_{2} m_{3} \left| U_{e2} \right|^{2} \left|U_{e3} \right|^{2} \cos(\alpha_{31} - \alpha_{21} ) \; .
\nonumber
\end{eqnarray}

The Lagrangian at high energy that describe the lepton sector of the SM  in the presence of the right-handed neurinos, $\nu_{R_{i}}$, is given by,
\begin{eqnarray}
\mathcal{L} & = & \overline{\ell}_{L_{i}} i \gamma^{\mu} \partial_{\mu} \ell_{L_{i}} + \overline{e}_{R_{i}} i \gamma^{\mu} \partial_{\mu} e_{R_{i}} + \overline{N}_{R_{i}} i \gamma^{\mu} \partial_{\mu} N_{R_{i}} 
\\
&&
+ f_{ij} \overline{e}_{R_{i}} \ell_{L_{j}} H^{\dagger} + h_{ij} \overline{N}_{R_{i}} \ell_{L_{j}}H - \frac{1}{2} M_{ij} N_{R_{i}}N_{R_{j}} + h.c. \; .
\nonumber
\end{eqnarray}
Without loose of generality, in the basis where $f_{ij}$ and $M_{ij}$ are diagonal, the Yukawa matrix $h_{ij}$ is  in general a complex 
matrix. For 3 families, $h$ has nine phases. Out of these nine phases, three can be absorbed into wave functions of $\ell_{L_{i}}$. Therefore, there are six physical phases remain. Furthermore, a real $h_{ij}$ can be diagonalized by a bi-unitary transformation, which is defined in terms of six mixing angles.  After integrating out the heavy Majorana neutrinos, the effective Lagrangian that describes the neutrino sector below the seesaw scale is,  
\begin{eqnarray}
\mathcal{L}_{eff} & = & \overline{\ell}_{L_{i}} i \gamma^{\mu} \partial_{\mu} \ell_{L_{i}} 
+ \overline{e}_{R_{i}} i \gamma^{\mu} \partial_{\mu} e_{R_{i}} 
+ f_{ii} \overline{e}_{R_{i}} \ell_{L_{i}}H^{\dagger}
\\
& & + \frac{1}{2} \sum_{k} h_{ik}^{T} h_{kj} \ell_{L_{i}} \ell_{L_{j}} \frac{H^{2}}{M_{k}} + h.c. \; .
\nonumber
\end{eqnarray} 
This leads to an effective neutrino Majorana  mass matrix whose parameters can be measured at the oscillation experiments. As 
Majorana mass matrix is symmetric, for three families, it has six independent complex elements and thus six complex phases. 
Out of these six phases, three of them can be absorbed into the wave functions of the charged leptons. Hence at low energy, there are only three physical phases and three mixing angles in the lepton sector. Going from high energy to low energy, the numbers of mixing angles and phases are thus reduced by half. Due to the presence of the additional mixing angles and complex phases in the heavy neutrino sector, it is generally not possible to connect leptogenesis with low energy CP violation.  However, in some specific models, such connection can be established. This will be discussed in more details in Sec.~\ref{sec:connect}.

\section{Standard Leptogenesis}\label{sec:stlpg}

\subsection{Standard Leptogenesis (Majorana Neutrinos)}

As mentioned in the previous section, baryon number violation arises naturally in many grand unified theories. In  the  GUT baryogenesis, the asymmetry is generated through the decays of heavy gauge bosons (denoted by ``V'' in the following) or leptoquarks (denoted by ``S'' in the following), which are particles that carry both $B$ and $L$ numbers. 
In GUTs based on $SU(5)$, the heavy gauge bosons or heavy leptoquarks have the following $B$-non-conserving decays, 
\begin{eqnarray}
V \rightarrow  \overline{\ell}_{L} u_{R}^{c}, \qquad & B = -1/3, \quad  & B-L = 2/3
\\
V \rightarrow  q_{L} d_{R}^{c}, \qquad & B = 2/3, \quad  & B-L = 2/3
\\
S \rightarrow  \overline{\ell}_{L}\overline{q}_{L}, \qquad & B = -1/3, \quad &  B-L = 2/3
\\
S \rightarrow  q_{L} q_{L}, \qquad & B = 2/3, \quad  & B-L = 2/3 \; .
\end{eqnarray}
Since $(B-L)$ is conserved, {\it i.e.} the heavy particles V and S both carry $(B-L)$ charges $2/3$, no $(B-L)$ can be generated dynamically. In addition, due to the sphaleron processes, $\left <B\right> = \left< B-L \right> = 0$. In $SO(10)$, $(B-L)$ is spontaneously broken, as it is a gauged subgroup of $SO(10)$. Heavy particles $X$ with $M_{X} < M_{B-L}$ can then generate a $(B-L)$ asymmetry through their decays. Nevertheless, for $M_{X} \sim M_{GUT} \sim 10^{15}$ GeV, the CP asymmetry  is highly suppressed. Furthermore, one also has to worry about the large reheating temperature $T_{RH} \sim M_{GUT}$ after the inflation, the realization of thermal equilibrium, and in supersymmetric case, the gravitino problem. These difficulties in GUT baryogenesis had led to a lot of interests in EW baryogenesis, which also has its own disadvantages as discussed in Sec.~\ref{sec:bg}. 

The recent advent of the evidence of neutrino masses from various neutrino oscillation experiments opens up a new possibility of generating the asymmetry through the decay of the heavy neutrinos~\cite{Fukugita:1986hr}. A particular attractive framework in which small neutrino masses can naturally arise is GUT based on $SO(10)$  (for a review, see, {\it i.e.} Ref.~\cite{Chen:2003zv}). $SO(10)$ GUT models accommodate the existence of RH neutrinos,
\begin{equation}
\psi(16) = (q_{L}, \; u_{R}^{c}, \; e_{R}^{c}, \; d_{R}^{c}, \; \ell_{L}, \; \nu_{R}^{c}) \; ,
\end{equation}
which is unified along with the fifteen known fermions of each family into a single 16-dimensional spinor representation. 
For hierarchical fermion masses, one easily has
\begin{equation}M_{N} \ll M_{B-L} \sim M_{GUT} \; ,
\end{equation}
where $N=\nu_{R} + \nu_{R}^{c}$ is a Majorana fermion.
The decays of the right-handed neutrino,
\begin{equation}
N \rightarrow \ell H, \quad N \rightarrow \overline{\ell} \, \overline{H} \; ,
\end{equation}
where $H$ is the SU(2) Higgs doublet, can lead to a lepton number asymmetry. After the sphaleron processes, the lepton number asymmetry is then converted into a baryon number asymmetry.  

The most general Lagrangian involving charged leptons and neutrinos is given by,
\begin{equation}
\mathcal{L}_{Y} = f_{ij} \overline{e}_{R_{i}} \ell_{L_{j}} H^{\dagger}+h_{ij} \overline{\nu}_{R_{i}} \ell_{L_{j}} H 
- \frac{1}{2} (M_{R})_{ij} \overline{\nu}_{R_{i}}^{c} \nu_{R_{j}}  + h.c. \; . 
\end{equation}
As the RH neutrinos are singlets under the SM gauge group, Majorana masses for the RH neutrinos are allowed by the gauge invariance. Upon the electroweak symmetry breaking, the SM Higgs doublet gets a VEV, $\left< H \right> = v$, 
 and the charged leptons and the neutrino Dirac masses, which are much smaller than the RH neutrino Majorana masses, are generated,
\begin{equation}
m_{\ell} = f v, \quad m_{D} = h v \ll M_{R} \; .
\end{equation}
The neutrino sector is therefore described by a $2\times 2$ seesaw matrix as,
\begin{equation}
\left(\begin{array}{cc}
0 & m_{D} \\
m_{D}^{T} & M_{R}
\end{array}\right) \;.
\end{equation}
Diagonalizing this $2\times 2$ seesaw matrix, the light and heavy neutrino mass eigenstates are obtained as,
\begin{equation}
\nu \simeq V_{\nu}^{T} \nu_{L} + V_{\nu}^{\ast} \nu_{L}^{c}, \quad 
N \simeq \nu_{R} + \nu_{R}^{c}
\end{equation}
with corresponding  masses
\begin{equation}
m_{\nu} \simeq -V_{\nu}^{T} m_{D}^{T} \frac{1}{M_{R}} m_{D} V_{\nu}, \quad 
m_{N} \simeq M_{R} \; .
\end{equation}
Here the unitary matrix $V_{\nu}$ is the diagonalization matrix of the neutrino Dirac matrix. 

At temperature $T < M_{R}$, RH neutrinos can generate a lepton number asymmetry by means of out-of-equilibrium decays. 
The sphaleron processes then convert $\Delta L$ into $\Delta B$.

\subsubsection{The Asymmetry}

\begin{figure}[t]
\vspace{0.3in}
\begin{tabular}{ccccc}
\includegraphics[scale=0.25]{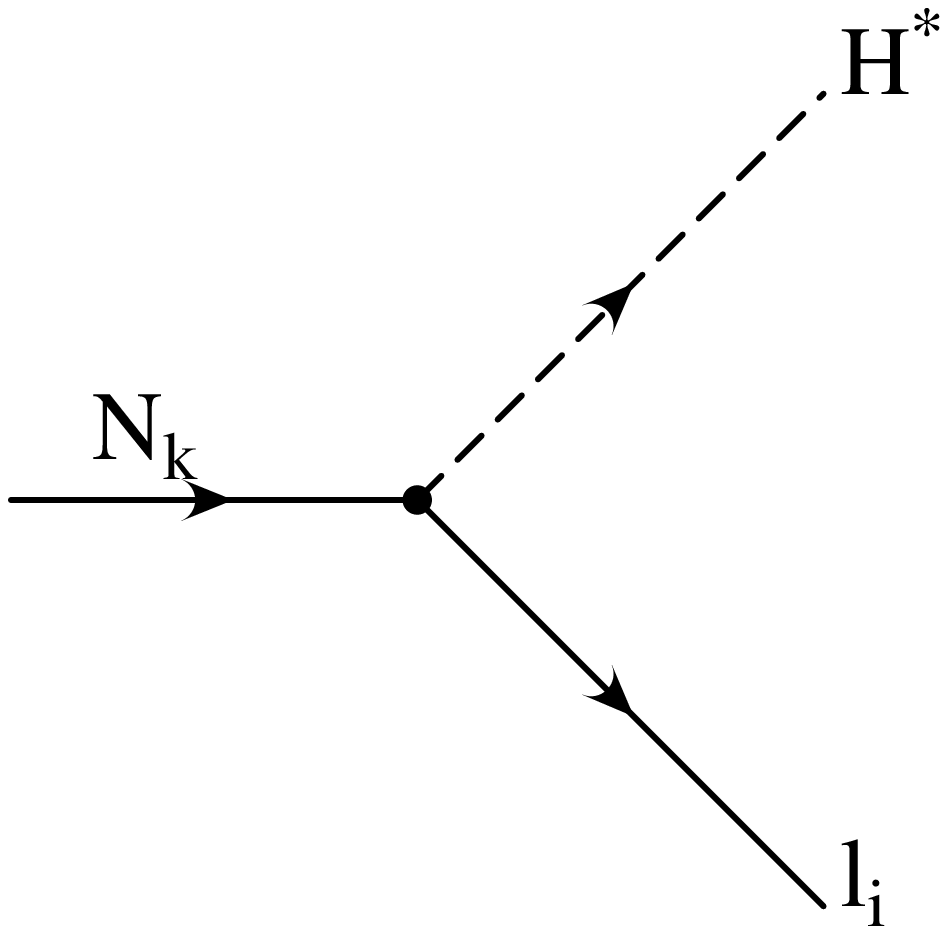} &
\hspace{0.2in} &
\includegraphics[scale=0.55]{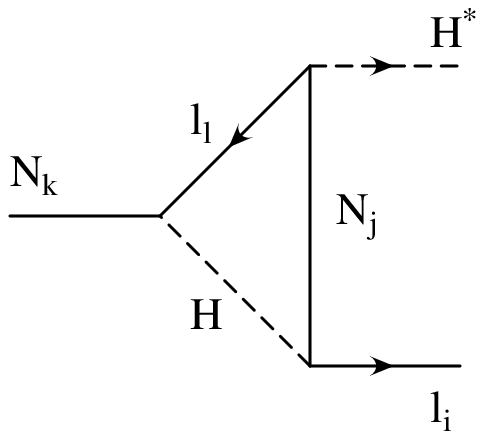} & 
\hspace{0.2in} & 
\includegraphics[scale=0.55]{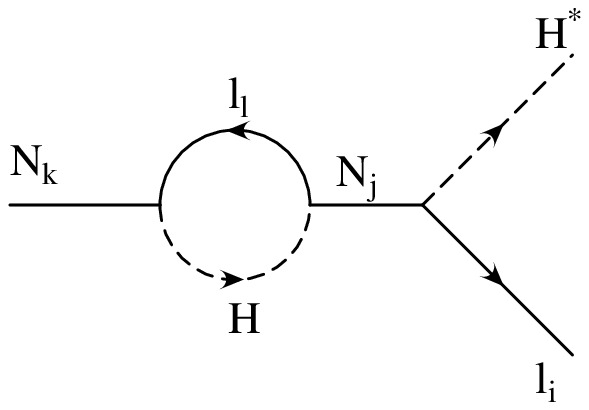} \\
(a) &   & (b) &  & (c)  \\
    \end{tabular}
    \vspace{0.1in}
    \caption{Diagrams in SM with RH neutrinos that contribute to the lepton number asymmetry through the decays of the RH neutrinos. The asymmetry is generated due to the interference of the tree-level diagram (a) and the one-loop vertex correction (b) and self-energy (c) diagrams.}
    \label{fig:lepg-std}
    \vspace{0.2in}
\end{figure}

At the tree level, the $i$-th RH neutrino decays into the Higgs doublet and the charged lepton doublet of $\alpha$ flavor, $N_{i} \rightarrow H + \ell_{\alpha}$, where $\alpha = (e, \; \mu, \; \tau)$. The total width of this decay is,
\begin{eqnarray}
\Gamma_{D_{i}} & = & \sum_{\alpha} \biggl[ \Gamma(N_{i} \rightarrow H + \ell_{\alpha}) + \Gamma(N_{i} \rightarrow \overline{H} + \overline{\ell}_{\alpha}) \biggr] \\
&= & \frac{1}{8\pi} (hh^{\dagger})_{ii} M_{i} \; . \nonumber
\end{eqnarray}
Suppose that the lepton number violating interactions of the lightest right-handed neutrino, $N_{1}$, wash out any lepton number asymmetry generated in the decay of $N_{2,3}$ at temperatures $T \gg M_{1}$. (For effects due to the decays of  $N_{2,3}$, see Ref.~\cite{Blanchet:2006dq}.)  
In this case with $N_{1}$ decay dominating, the final asymmetry only depends on the dynamics of $N_{1}$.  
The out-of-equilibrium condition requires that the total width for $N_{1}$ decay, $\Gamma_{D_{1}}$, to be smaller compared to the expansion rate of the Universe at temperature $T = M_{1}$,
\begin{equation}
\Gamma_{D_{1}} < H \bigg|_{T=M_{1}} \; .
\end{equation}
That is, the heavy neutrinos are not able to follow the rapid change of the equilibrium particle distribution, once the temperature dropped below the mass $M_{1}$. Eventually, heavy neutrinos will decay, and a lepton asymmetry is generated due to the CP asymmetry that arises through the interference of the tree level and one-loop diagrams, as shown in Fig.~\ref{fig:lepg-std}, 
\begin{eqnarray}
\epsilon_{1} & = & \frac{\sum_{\alpha} \bigl[\Gamma(N_{1} \rightarrow \ell_{\alpha} H) - \Gamma(N_{1} \rightarrow \overline{\ell}_{\alpha} \, \overline{H}) \bigr]}
{\sum_{\alpha} \bigl[\Gamma(N_{1} \rightarrow \ell_{\alpha} H) + \Gamma(N_{1} \rightarrow \overline{\ell}_{\alpha} \, \overline{H})\bigr]} \label{eq:e1}
\\
& \simeq & 
\frac{1}{8\pi} \frac{1}{(h_{\nu}h_{\nu})_{11}} \sum_{i=2,3} \mbox{Im}\bigg\{(h_{\nu}h_{\nu}^{\dagger})_{1i}^{2}\bigg\}
\cdot \biggl[ f\biggl( \frac{M_{i}^{2}}{M_{1}^{2}}\biggr) 
+ g\biggl( \frac{M_{i}^{2}}{M_{1}^{2}}\biggr)\biggr] \; .
\nonumber
\end{eqnarray}
In Fig.~\ref{fig:lepg-std}, the diagram (b) is the one-lop vertex correction, which gives the term, $f(x)$, in Eq.~\ref{eq:e1} after carrying out the loop integration,
\begin{equation}
f(x) = \sqrt{x} \biggl[ 1-(1+x) \ln \biggl(\frac{1+x}{x}\biggr) \biggr] \; .
\end{equation}
Diagram (c) is the one-loop self-energy. For $|M_{i} - M_{1}| \gg | \Gamma_{i} - \Gamma_{1} |$, the self-energy diagram gives the term 
\begin{equation}
g(x) = \frac{\sqrt{x}}{1-x} \; ,
\end{equation}
in Eq.~\ref{eq:e1}. For hierarchical RH neutrino masses, $M_{1} \ll M_{2}, \; M_{3}$, the asymmetry is then given by, 
\begin{equation}
\epsilon_{1} \simeq -\frac{3}{8\pi} \frac{1}{(h_{\nu}h_{\nu}^{\dagger})_{11}} \sum_{i=2,3} 
\mbox{Im} \bigg\{ (h_{\nu}h_{\nu}^{\dagger})_{1i}^{2}\bigg\} \frac{M_{1}}{M_{i}} \; .
\end{equation}
Note that when $N_{k}$ and $N_{j}$ in the self-energy diagram (c) have near degenerate masses, there can be resonant enhancement in the contributions from the self-energy diagram to the asymmetry. Such resonant effect can allow $M_{1}$ to be much lower while still generating sufficient amount of the lepton number asymmetry. This will be discussed in Sec.~\ref{sec:res}.


To prevent the generated asymmetry given in Eq.~\ref{eq:e1} from being washed out by the inverse decay and scattering processes, the decay of the RH neutrinos has to be out-of-equilibrium. In other words, the condition
\begin{equation}
r \equiv \frac{\Gamma_{1}}{H|_{T=M_{1}}}=\frac{M_{pl}}{(1.7)(32\pi)\sqrt{g_{\ast}}}\frac{(h_{\nu}h_{\nu}^{\dagger})_{11}}{M_{1}} < 1\; ,
\end{equation}
has to be satisfied. 
This leads to the following constraint on the effective light neutrino mass
\begin{equation}\label{eq:mtilde}
\widetilde{m}_{1} \equiv (h_{\nu}h_{\nu}^{\dagger})_{11} \frac{v^{2}}{M_{1}} \simeq 
4 \sqrt{g_{\ast}} \frac{v^{2}}{M_{pl}} \frac{\Gamma_{D_{1}}}{H} \bigg|_{T=M_{1}} < 10^{-3} \; \mbox{eV} \; ,
\end{equation}
where  $g_{\ast}$ is the number of relativistic degrees of freedom. For SM, $g_{\ast} \simeq 106.75$, while for MSSM, $g_{\ast} \simeq 228.75$.
The wash-out effect is parameterized by the coefficient $\kappa$, and the final amount of lepton asymmetry is given by, 
\begin{equation}
Y_{L} \equiv \frac{n_{L} - \overline{n}_{L}}{s} = \kappa \frac{\epsilon_{1}}{g_{\ast}} \; ,
\end{equation}
where  $\kappa$ parameterizes the amount of wash-out due to the inverse decays and scattering processes. 
The amount of wash-out depends on the size of the parameter $r$: 
\begin{enumerate}
\item If $r \ll1$ for decay temperature $T_{D} \lesssim M_{X}$, the inverse decay and 2-2 scattering are impotent.  In this case, the inverse decay width is given by,
\begin{equation}
\frac{\Gamma_{ID}}{H}  \sim  \biggl( \frac{M_{X}}{T} \biggr)^{3/2} e^{-M_{X}/T} \cdot r \; ,
\end{equation}
while the width for the scattering processes is,
\begin{equation}
\frac{\Gamma_{S}}{H}  \sim  \alpha \biggl(\frac{T}{M_{X}}\biggr)^{5} \cdot r \; .
\end{equation}
Thus the inverse decays and scattering processes can be safely ignored, and 
the asymmetry $\Delta B$ produced by decays is not destroyed by the asymmetry $-\Delta B$ produced in inverse decays and scatterings. 
At $T \simeq T_{D}$, the number density of the heavy particles $X$ has thermal distribution, $n_{X} \simeq n_{\overline{X}} \simeq n_{\gamma}$. 
Thus the net baryon neumber density produced by out-of-equilibrium decays is
\begin{equation}
n_{L} = \epsilon_{1} \cdot n_{X} \simeq \epsilon_{1} \cdot n_{\gamma} \; .
\end{equation}

\item For $r \gg1$, the abundance of X and $\overline{X}$ follows the equilibrium values, and there is no departure from thermal equilibrium. As a result, no lepton number may evolve, and the net lepton asymmetry vanishes, 
\begin{equation}
\frac{n_{\ell}-n_{\overline{\ell}}}{dt} + 3 H (n_{\ell}-n_{\overline{\ell}}) = \Delta \gamma^{eq} = 0 \; .
\end{equation}

\end{enumerate}

In general, for $1 < r < 10$, there could still be sizable asymmetry. The wash out effects due to inverse decay and lepton number violating scattering processes together with the time evolution of the system is then accounted for by the factor $\kappa$, which is obtained by solving the Bolzmann equations for the system (see next section). An approximation is given by~\cite{Kolb:1988aj}, 
\begin{eqnarray}
10^{6} \lesssim r:  & \kappa = (0.1r)^{1/2} e^{-\frac{4}{3}(0.1)^{1/4}} \quad (<10^{-7}) \\
10 \lesssim r \lesssim 10^{6}: & \kappa = \frac{0.3}{r (\ln r)^{0.8}} \quad (10^{-2} \sim 10^{-7})\\
0 \lesssim r \lesssim 10: & \kappa = \frac{1}{2\sqrt{r^{2} + 9}} \quad (10^{-1} \sim 10^{-2}) \; .
\end{eqnarray}

The EW sphaleron effects then convert $Y_{L}$ into $Y_{B}$, 
\begin{equation}
Y_{B} \equiv \frac{n_{B} - n_{\overline{B}}}{s} = c Y_{B-L} = \frac{c}{c-1} Y_{L} \; ,
\end{equation}
where $c$ is the conversion factor derived in Sec.~\ref{sec:convert}.

\subsubsection{Boltzmann Equations}\label{sec:boltz}

As the decays of RH neutrinos are out-of-equilibrium processes, they are generally treated by Boltzmann equations. 
Main processes in the thermal bath that are relevant for leptogenesis include,
\begin{figure}[t]
\centerline{
\vspace{0.2in}
\begin{tabular}{ccc}
\includegraphics[scale=0.25]{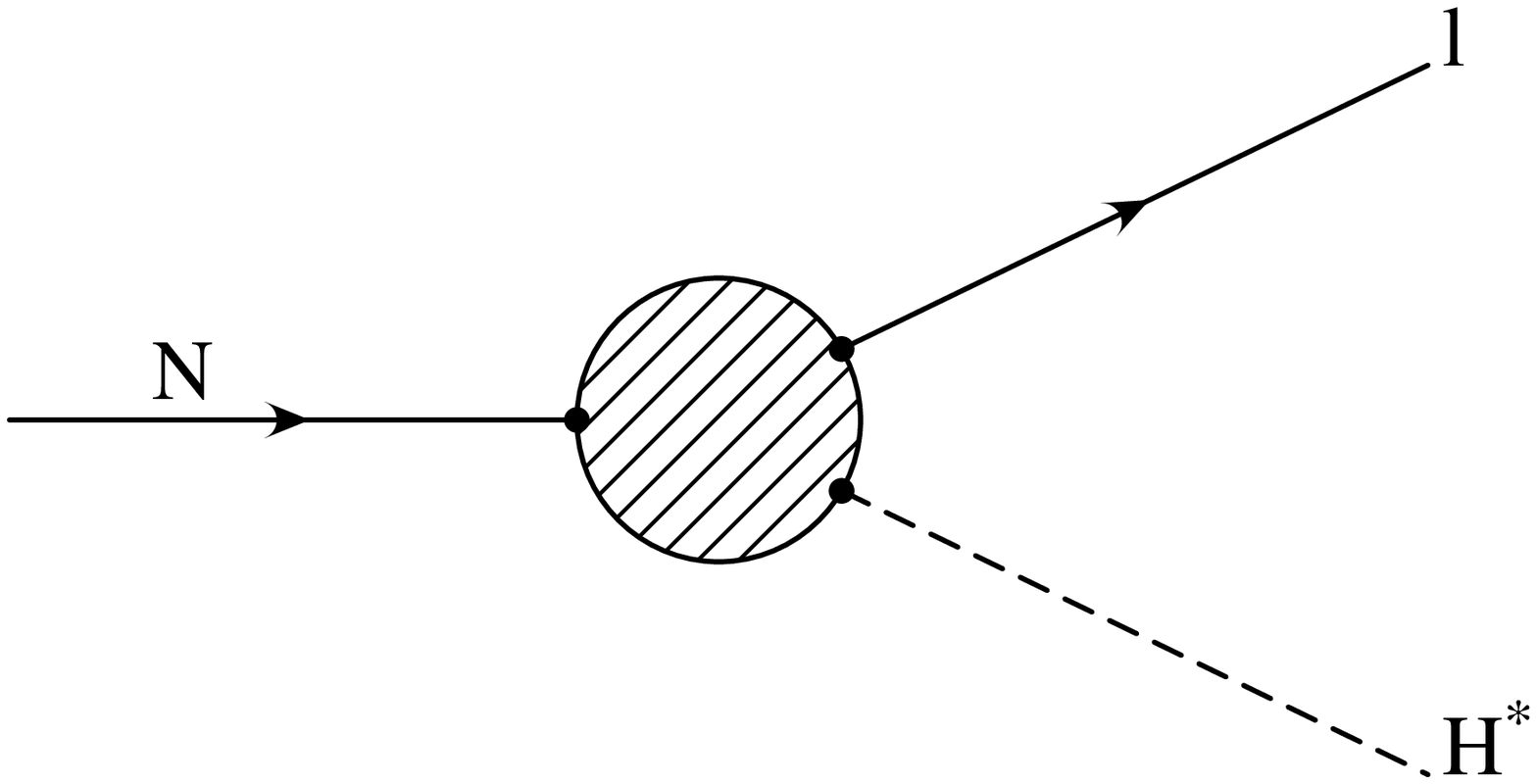} &
\hspace{0.2in} &
\includegraphics[scale=0.25]{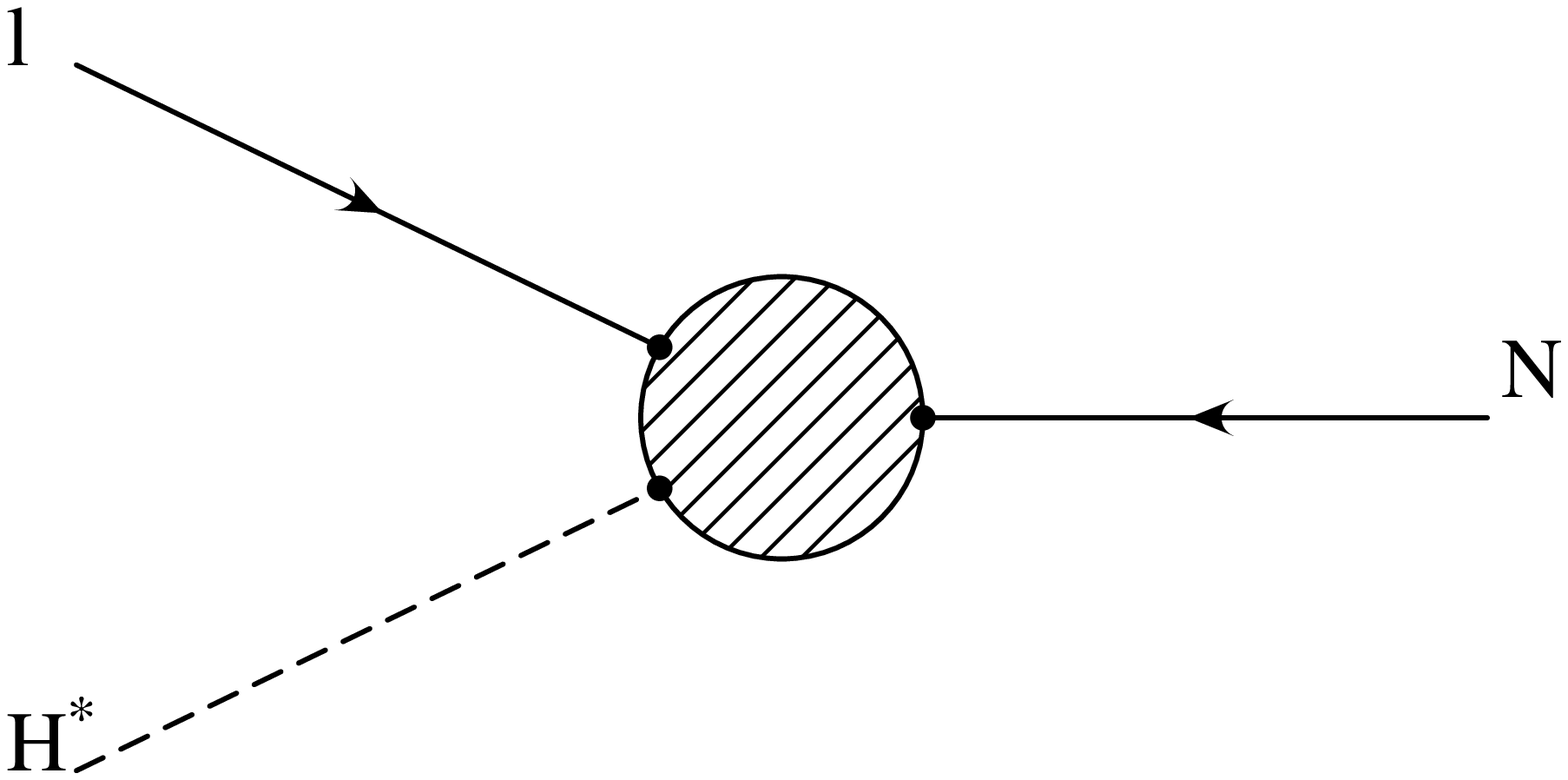} \\
(a) &   & (b)   \\
    \end{tabular}}
 \caption{Decay and inverse decay processes in the thermal bath.}
 \label{fig:Ndecay1}
 \end{figure}
\begin{figure}[h!]
\centerline{
\vspace{0.2in}
\begin{tabular}{ccc}
\includegraphics[scale=0.25]{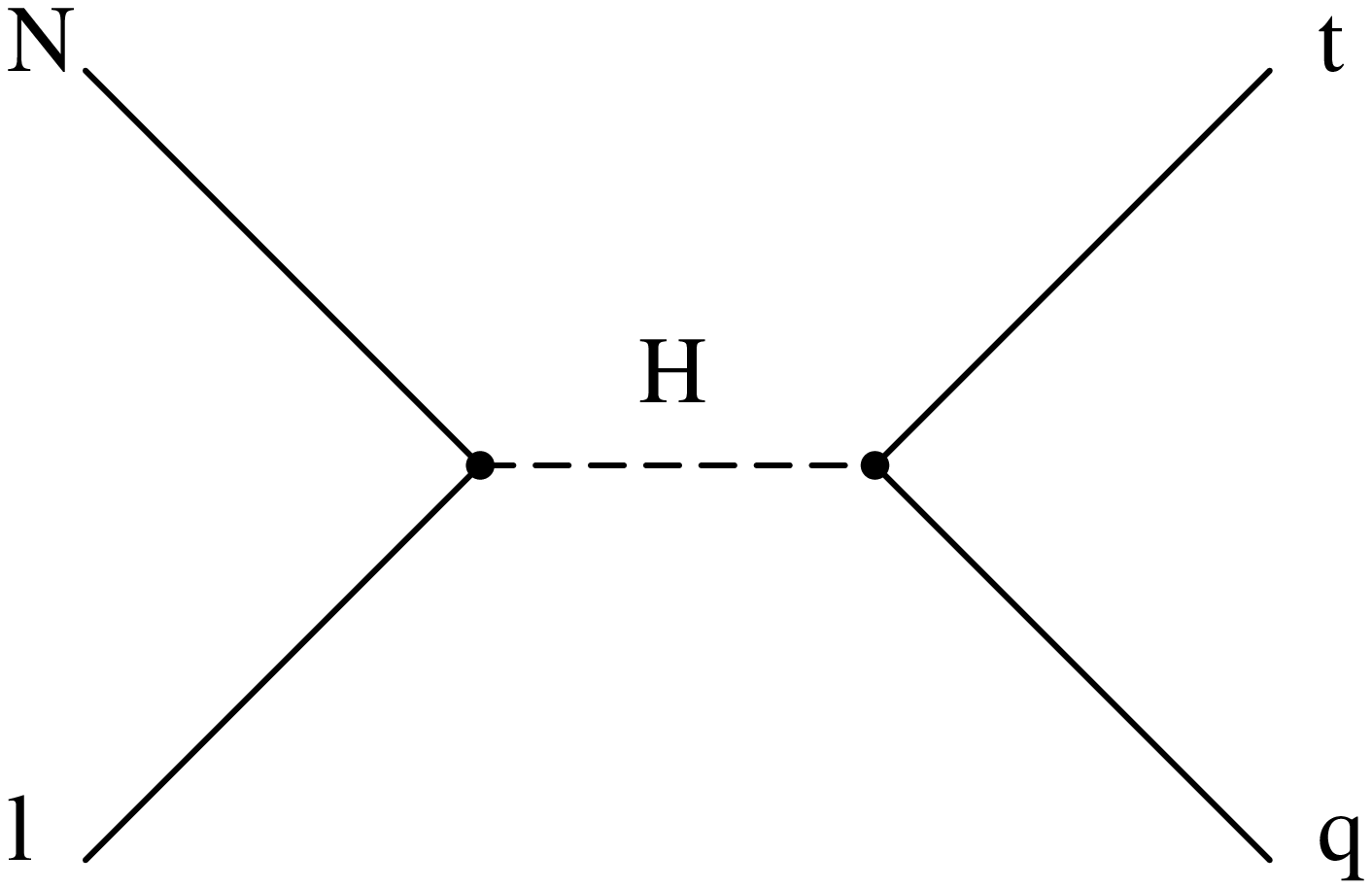} &
\hspace{0.2in} &
\includegraphics[scale=0.25]{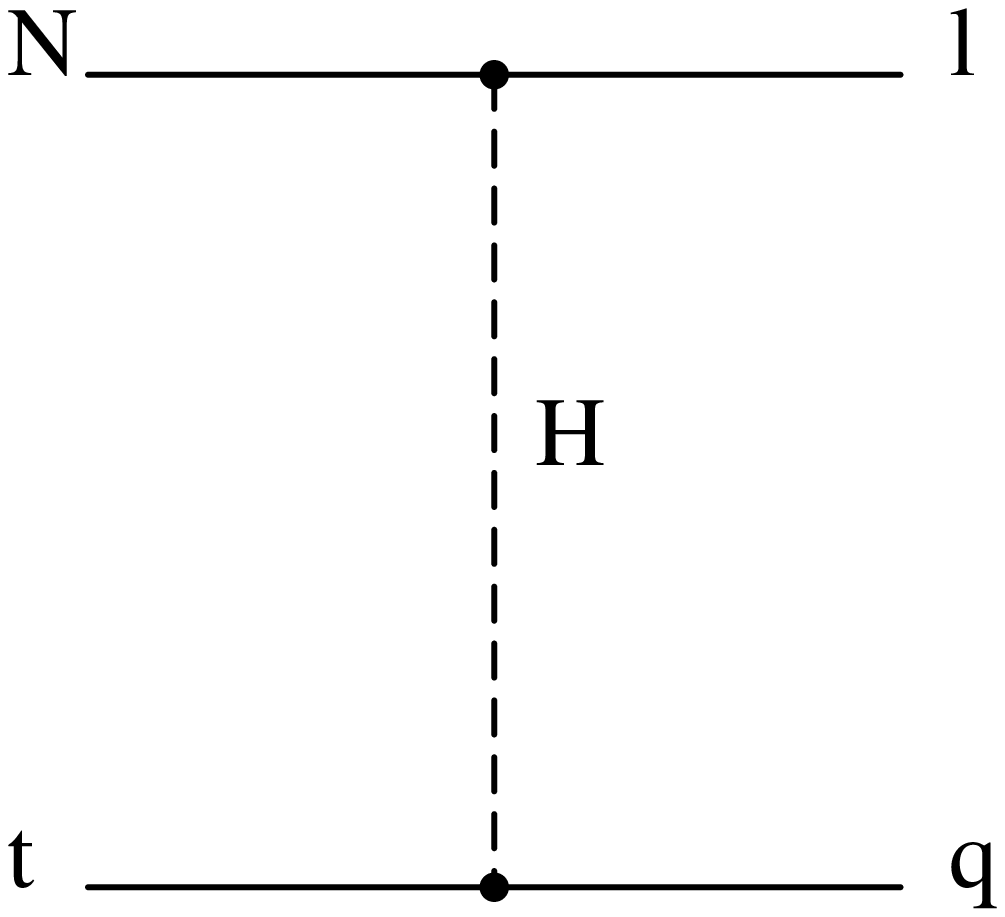} \\
&&\\
(a) &   & (b)   \\
    \end{tabular}}
 \caption{The $\Delta L = 1$  scattering processes in the thermal bath.}
 \label{fig:scatt-L1}
 \end{figure}
\begin{figure}[h!]
\centerline{
\vspace{0.2in}
\begin{tabular}{ccccc}
\includegraphics[scale=0.25]{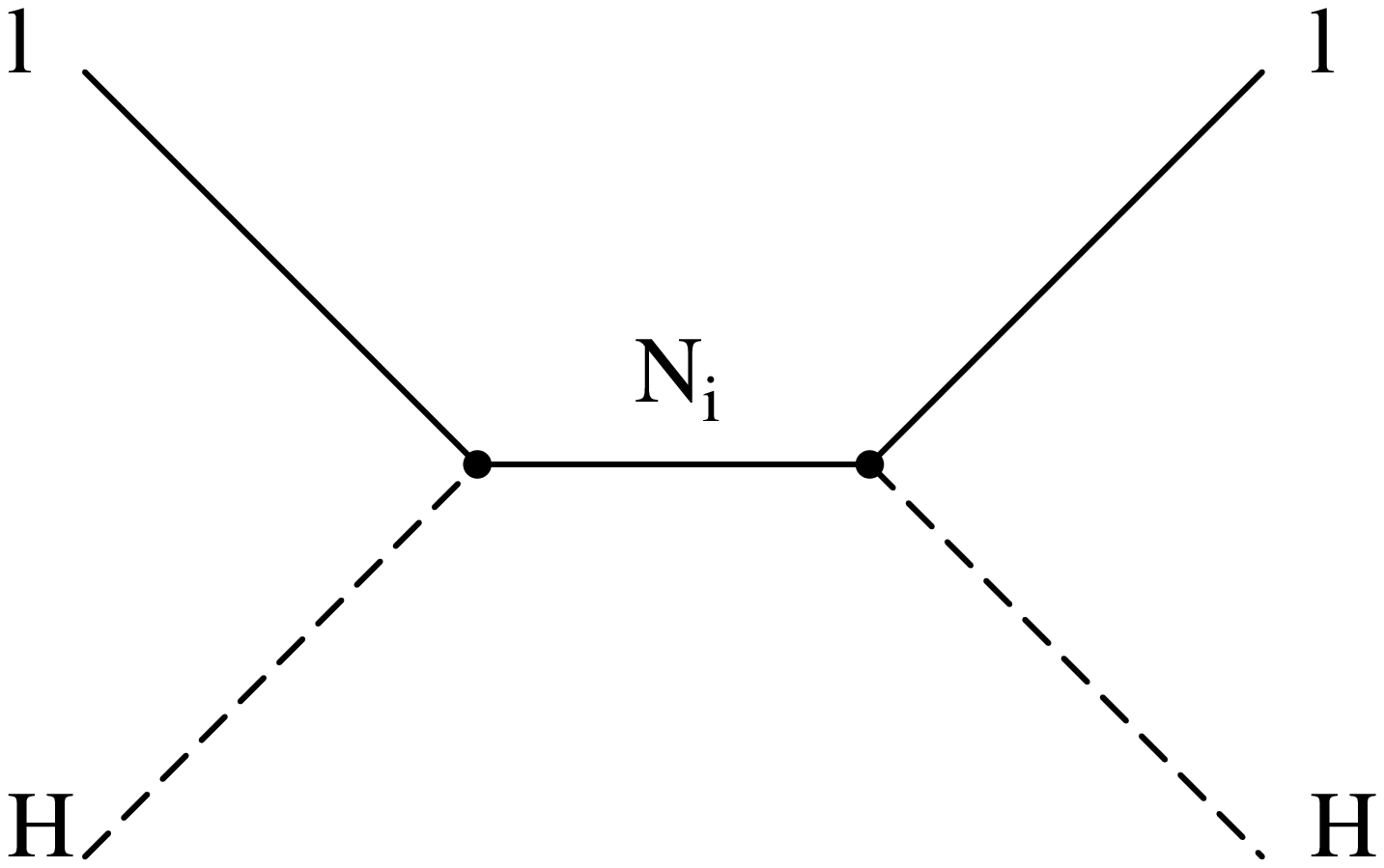} &
\hspace{0.2in} &
\includegraphics[scale=0.25]{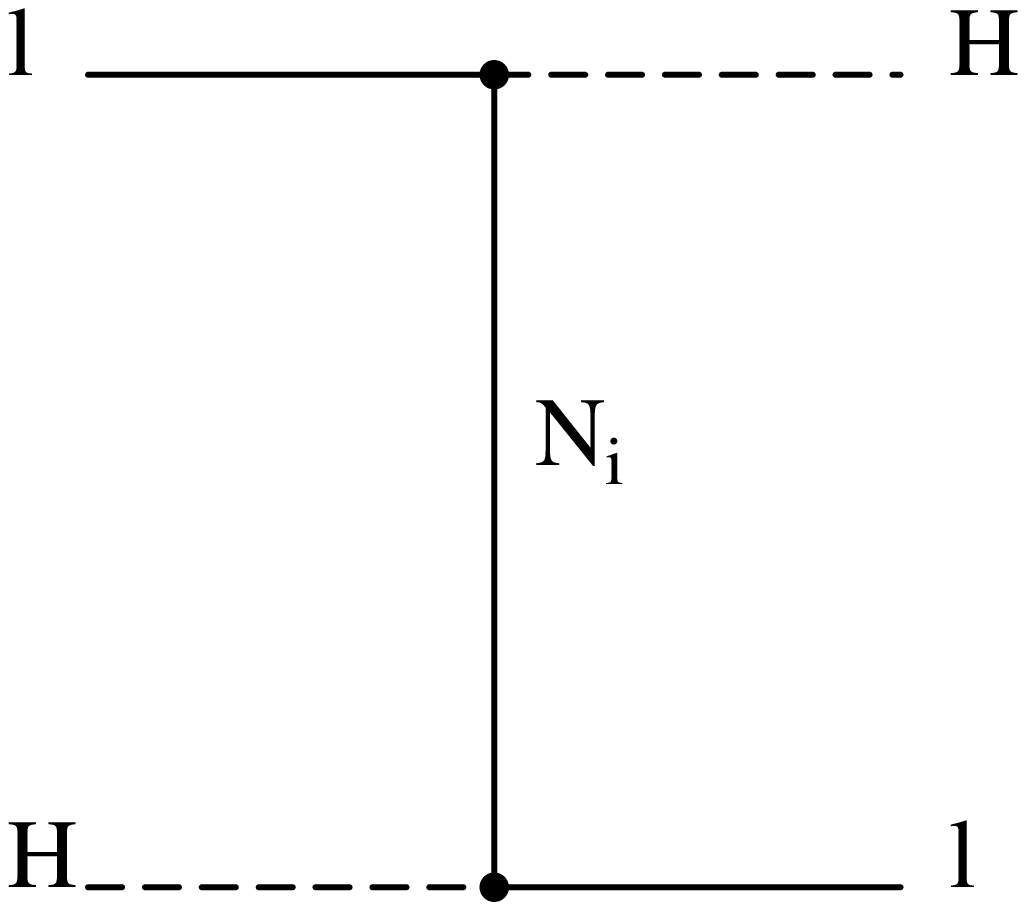} &
\hspace{0.2in} &
\includegraphics[scale=0.25]{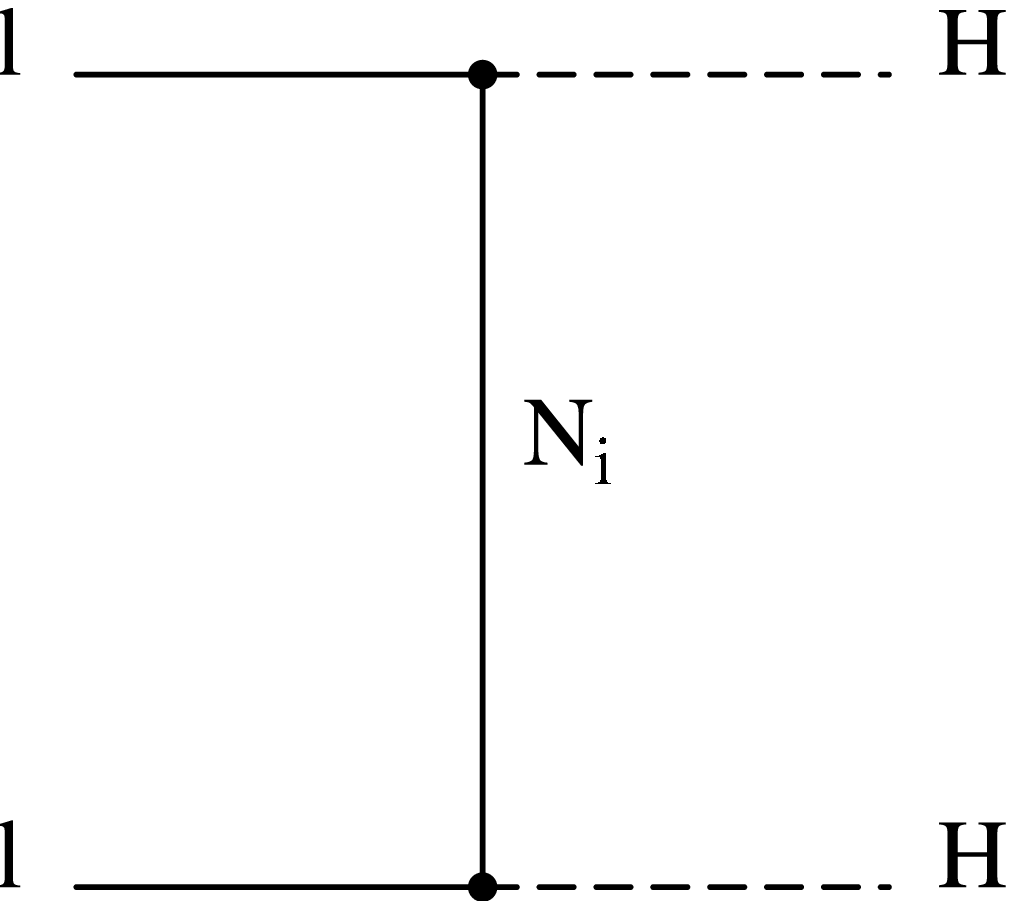} \\
&&\\
(a) &   & (b) &  &(c)  \\
    \end{tabular}}
 \caption{The $\Delta L = 2$  scattering processes in the thermal bath.}
 \label{fig:scatt-L2}
 \vspace{0.2in}
 \end{figure}

\begin{enumerate}

\item decay of N  (Fig.~\ref{fig:Ndecay1} (a)):   
\begin{equation}
N\rightarrow \ell + H, \qquad N\rightarrow \overline{\ell} + \overline{H}
\end{equation}

\item inverse decay of $N$  (Fig.~\ref{fig:Ndecay1} (b)): 
\begin{equation}
\ell + H \rightarrow N, \qquad  \overline{\ell} + \overline{H} \rightarrow N
\end{equation}

\item 2-2 scattering: These include the following $\Delta L = 1$ scattering processes (Fig.~\ref{fig:scatt-L1}), 
\begin{eqnarray}
 \mbox{[s-channel]} : & ~~~~~N_{1} \, \ell   \leftrightarrow   t \, \overline{q}
\quad , \quad
N_{1} \, \overline{\ell}  \leftrightarrow   t \, \overline{q}
\\
\mbox{[t-channel]}: & ~~~~~N_{1} t \leftrightarrow  \overline{\ell} \, q 
\quad, \quad
N_{1} \, \overline{t}  \leftrightarrow  \ell \,  \overline{q} 
\end{eqnarray}
and $\Delta L = 2$ scattering processes (Fig.~\ref{fig:scatt-L2}),
\begin{eqnarray}
\ell H   \leftrightarrow  \overline{\ell} \,  \overline{H} 
\quad , \quad
\ell \ell  \leftrightarrow  \overline{H} \, \overline{H}, \quad \overline{\ell} \,  \overline{\ell} \leftrightarrow  H \, H  \; .
\end{eqnarray}
\end{enumerate}
Basically, at temperatures $T \gtrsim M_{1}$, these $\Delta L = 1$ and $\Delta L = 2$ processes have to be strong enough to keep $N_{1}$ in equilibrium. Yet at temperature $T \lesssim M_{1}$, these processes have to be weak enough to allow $N_{1}$ to generate an asymmetry.

The Boltzmann equations that govern the evolutions of the RH neutrino number density and $B-L$ number density are given by~\cite{Buchmuller:2002rq},
\begin{eqnarray}
\frac{dN_{N_{1}}}{dz} & = & -(D+S)(N_{N_{1}}-N_{N_{1}}^{eq}) \label{eq:btz1}
\\
\frac{dN_{B-L}}{dz} & = & -\epsilon_{1} D(N_{N_{1}}-N_{N_{1}}^{eq}) - WN_{B-L} \; , \label{eq:btz2}
\end{eqnarray}
where 
\begin{equation}
(D,S,W) \equiv \frac{(\Gamma_{D},\Gamma_{S},\Gamma_{W})}{Hz}, \quad z=\frac{M_{1}}{T} \; .
\end{equation}
\begin{figure}[b!]
\centerline{
\psfig{file=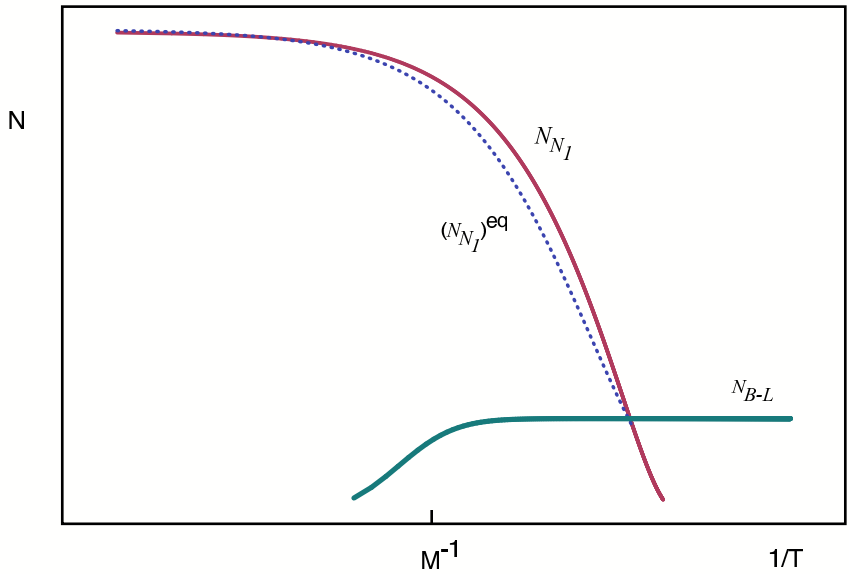,width=12.0cm}\hspace{1.2cm}}    
 \caption{Generic behavior of the solutions to Boltzmann equations. Here the functions $N_{N_{1}}$ (red solid curve) and $N_{B-L}$ (green solid 
 curve) are solutions to Eq.~\ref{eq:btz1} and \ref{eq:btz2}. The function $(N_{N_{1}})^{eq}$ (blue dotted curve) is the equilibrium particle distribution.}
    \label{fig:btz}
\end{figure}
Here $\Gamma_{D}$ includes both decay and inverse decay, $\Gamma_{S}$ includes $\Delta L=1$ scattering processes
and $\Gamma_{W}$ includes inverse decay and  $\Delta L=1$, $\Delta L=2$ scattering processes. The $N_{1}$ abundance is affected by the decay, inverse decay and the $\Delta L = 1$ scattering processes. It is manifest in Eq.~\ref{eq:btz2} that  the $N_{1}$ decay is the source for $(B-L)$, while the inverse decay and the $\Delta L = 1, \, 2$ scattering processes wash out the asymmetry. The generic behavior of the solutions to the Boltzmann equations is shown in Fig.~\ref{fig:btz}.

\subsubsection{Bounds on Neutrino Masses}\label{sec:bounds}

In the case with strongly hierarchical right-handed neutrino masses, when the asymmetry $\epsilon_{1}$ due to the decay of the lightest right-handed neutrino, $N_{1}$, contribute dominantly to the total asymmetry, leptogenesis becomes very 
predictive~\cite{Buchmuller:2002rq,Buchmuller:2005eh,Nir:2007zq}, provided that $N_{1}$ decays at temperature $T \gtrsim 10^{12}$ GeV. In particular, various bounds on the neutrino masses can be obtained. 

For strongly hierarchyical masses, $M_{1}/M_{2} \ll 1$, there is an upper bound on $\epsilon_{1}$~\cite{Davidson:2002qv}, called the ``Davidson-Ibarra'' bound, 
\begin{equation}
|\epsilon_{1}| \le \frac{3}{16\pi} \frac{M_{1}(m_{3}-m_{2})}{v^{2}} \equiv \epsilon_{1}^{DI} \; ,
\end{equation}
which is obtained by expanding $\epsilon_{1}$ to leading order in $M_{1}/M_{2}$. 
Becuase $|m_{3} - m_{2}| \le \sqrt{\Delta m_{32}^{2}} \sim 0.05$ eV, a lower bound on $M_{1}$ then follows,
\begin{equation}\label{eq:m1bound}
M_{1} \ge 2 \times 10^{9} \; \mbox{GeV} \; .
\end{equation} 
This bound in turn implies a lower bound on the reheating temperature, $T_{RH}$, and is in conflict with the upper bound from gravitino over production constraints if supersymmetry is incorporated. We will come back to this in Sec.~\ref{sec:gravitino}. One should note that, in the presence of  degenerate light neutrinos, the leading terms in an expansion of $\epsilon_{1}$ in $M_{1}/M_{2}$ and $M_{1}/M_{3}$ vanish. However, the next to leading order terms do not vanish and in this case one has~\cite{Hambye:2003rt},
\begin{equation}
|\epsilon_{1}| \lesssim \mbox{Max}\biggl( \epsilon^{DI}, \frac{M_{3}^{3}}{M_{1}M_{2}^{2}}\biggr) \; .
\end{equation}

By requiring that there is no substantial washout effects, bounds on light neutrino masses can be derived. To have significant amount of baryon asymmetry, the effective mass $\widetilde{m}_{1}$ defined in Eq.~\ref{eq:mtilde} cannot be too large. Generally $\widetilde{m}_{1} \lesssim 0.1 - 0.2$ is required. As the mass of the lightest active neutrino $m_{1} \lesssim \widetilde{m}_{1}$, an upper bound on $m_{1}$ thus ensues. By further requiring the $\Delta L = 2$ washout effects be consistent with successful leptogenesis impose a bound on,
\begin{equation}
\sqrt{(m_{1}^{2} + m_{2}^{2} + m_{3}^{2})} \lesssim (0.1 - 0.2) \; \mbox{eV} \; ,
\end{equation}
which is of the same order as the bound on $\widetilde{m}_{1}$. From these bounds, the absolute mass scale of neutrino masses is thus known up to a factor of $\sim 3$ to be in the range, $0.05 \lesssim m_{3} \lesssim 0.15$ eV~\cite{Nir:2007zq}, if the observed baryonic asymmetry indeed originates from leptogenesis through the decay of $N_{1}$.

\subsection{Dirac Leptogenesis}

In the standard leptogenesis discussed in the previous section, neutrinos acquire their masses through the seesaw mechanism.  The decays of the heavy right-handed neutrinos produce a non-zero lepton number asymmetry, $\Delta L \ne 0$. The electroweak sphaleron effects then convert  $\Delta L$ partially into $\Delta B$. This standard scenario relies crucially on the violation of lepton number, which is due to the presence of the heavy Majorana masses for the right-handed neutrinos. 

It was pointed out ~\cite{Dick:1999je} that leptogenesis can be implemented even in the case when neutrinos are Dirac fermions which acquire small masses through highly suppressed Yukawa couplings without violating lepton number. The realization of this depends critically on the following three characteristics of the sphaleron effects: ({\it i}) only the left-handed particles couple to the sphalerons; ({\it ii}) the sphalerons change ($B+L$) but not ($B-L$); ({\it iii}) the sphaleron effects are in equilibrium for $T \gtrsim T_{EW}$. 

As the sphelarons couple only to the left-handed fermions, one may speculate that as long as the lepton number stored in the right-handed fermions can survive below the electroweak phase transition, a net lepton number may be generated even with $L=0$ initially. The Yukawa couplings of the SM quarks and leptons to the Higgs boson lead to rapid left-right equilibration so that as the sphaleron effects deplete the left-handed $(B+L)$, the right-handed $(B+L)$ is converted to fill the void and therefore it is also depleted.  So with $B=L=0$ initially, no baryon asymmetry can be generated for the SM quarks and leptons. For the neutrinos, on the other hand, the left-right equilibration can occur at a much longer time scale compared to the electroweak epoch when the sphaleron washout is in effect.  
\begin{figure}[t!]
\centerline{
\includegraphics{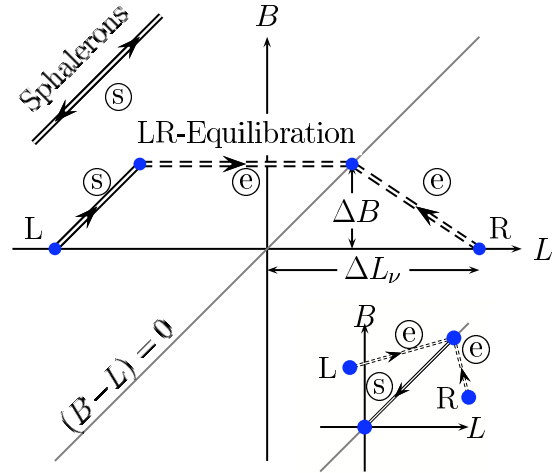}}    
 \caption{With sufficiently small Yukawa couplings, the left-right equilibration occurs at a much later time, well below the electroweak phase transition temperature. It is therefore possible to generate a non-zero baryon number even if $B=L=0$ initially.  
 For the SM particles, as shown in the insert for comparison, the left-right equilibration takes place completely before or during the sphaleron processes. Thus no net baryon number can be generated if $B-L=0$ initially. Figure taken from Ref~\cite{Dick:1999je}.}
    \label{fig:diraclep}
\end{figure}
The left-right conversion for the neutrinos involves the Dirac Yukawa couplings,  $\lambda \overline{\ell}_{L} H \nu_{R}$, where $\lambda$ is the Yukawa coupling constant, and the rate for these conversion processes scales as,
\begin{equation}
\Gamma_{LR} \sim \lambda^{2} T \; .
\end{equation}
For the left-right  conversion not to be in equilibrium at temperatures above some critical temperature $T_{eq}$, requires that
\begin{equation}
\Gamma_{LR} \lesssim H \; , \quad  \mbox{for} \quad T > T_{eq} \; ,
\end{equation}
where the Hubble constant scales as,
\begin{equation}
H \sim \frac{T^{2}}{M_{\mbox{\tiny Pl}}}  \; .
\end{equation}
Hence the left-right equilibration can occur at a much later time, $T \lesssim T_{eq} \ll T_{EW}$, provided, 
\begin{equation}
\lambda^{2} \lesssim \frac{T_{eq}}{M_{\mbox{\tiny Pl}}}  \ll \frac{T_{EW}}{M_{\mbox{\tiny Pl}}} \; .
\end{equation}
With $M_{\mbox{\tiny Pl}} \sim 10^{19}$ GeV and $T_{EW} \sim 10^{2}$ GeV, this condition then translates into
\begin{equation}
\lambda < 10^{-(8\sim9)} \; .
\end{equation}
 Thus for neutrino Dirac masses  $m_{D} < 10$ keV, which is consistent with all experimental observations, the left-right equilibration does not occur until the temperature of the Universe drops to much below the temperature of the electroweak phase transition, and the lepton number stored in the right-handed neutrinos can then survive the wash-out due to the sphalerons~\cite{Dick:1999je}. 
 
Once we accept this, the Dirac leptogenesis then works as follows. Suppose that some processes initially produce a negative lepton number ($\Delta L_{L}$), which is stored in the left-handed neutrinos, and a positive lepton number ($\Delta L_{R}$), which is  stored in the right-handed neutrinos. Because sphalerons only couple to the left-handed particles, part of the negative lepton number stored in left-handed neutrinos get converted into a positive baryon number by the electroweak anomaly. This negative lepton number $\Delta L_{L}$ with reduced magnitude eventually equilibrates with the positive lepton number, $\Delta L_{R}$ when the temperature of the Universe drops to $T \ll T_{EW}$. Because the equilibrating processes conserve both the baryon number $B$ and the lepton number $L$ separately, they result in a Universe with a total positive baryon number and a total positive lepton number. And hence a net baryon number can be generated even with $B=L=0$ initially. 

Such small neutrino Dirac Yukawa couplings required to implement Dirac leptogensis are realized in  a SUSY model proposed in Ref.~\cite{Murayama:2002je}.

\begin{figure}[b!]
\centerline{\psfig{file=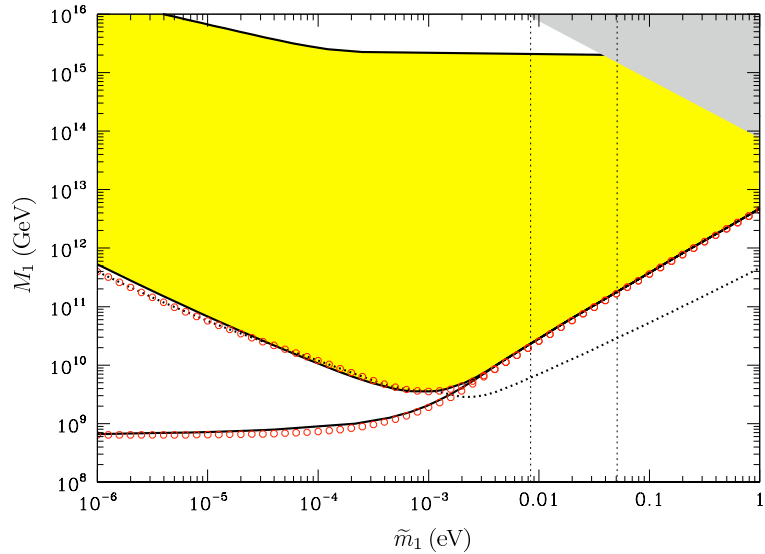,width=10.0cm}}
\caption{Lower bound on the lightest RH neutrino mass, $M_{1}$ (circles) and the initial temperature, $T_{i}$ (dotted line), for $m_{1} = 0$ and $\eta_{B}^{CMB} = 6 \times 10^{-10}$. The red circles (solid lines) denote the analytical (numerical) results. The vertical dashed lines indicate the range $(\sqrt{\Delta m_{\mbox{\tiny sol}}^{2}}, \sqrt{\Delta m_{\mbox{\tiny atm}}^{2}})$. 
Figure taken from Ref. \cite{Buchmuller:2004nz}.}
\label{fig:mlimit}
\end{figure}

\subsection{Gravitino Problem}\label{sec:gravitino}

For leptogenesis to be effective,  as shown in Sec.~\ref{sec:bounds}, the mass of the lightest RH neutrino has to be $M_{1} > 2 \times  10^{9}$ GeV.  Fig.~\ref{fig:mlimit} shows the lower bound on the lightest RH neutrino mass as a function of the low energy effective lightest neutrino mass, $\widetilde{m}_{1}$~\cite{Buchmuller:2002jk,Buchmuller:2004nz}. 
If RH neutrinos are produced thermally, the reheating temperature has to be greater than the right-handed neutrino mass, $T_{RH} > M_{R}$. This thus implies that $T_{RH} > 2 \times 10^{9}$ GeV, in order to generate sufficient baryon number asymmetry. Such a high reheating temperature is problematic as it could lead to overproduction of light states, such as gravitinos~\cite{Khlopov:1984pf,Giudice:1999am}. If gravitinos are stable (i.e. LSP), WMAP constraint on DM leads to stringent bound on gluino mass for any given gravitino mass $m_{3/2}$ and reheating temperature $T_{RH}$. (Bounds on other gaugino masses can also be obtained as discussed in~\cite{Pradler:2006qh}.) On the other hand, if gravitinos are unstable, it has long lifetime and can decay during and after the BBN, and may have the following three effects on BBN~\cite{Buchmuller:2005eh}: 
\begin{enumerate}
\item These decays can speed up cosmic expansion, and increase the neutron to proton ratio and thus the $^{4}$He abundance; 
\item Radiation decay of gravitinos, $\psi \rightarrow \gamma + \tilde{\gamma}$, increases the photon density and thus reduces the $n_{B}/n_{\gamma}$ ratio; 
\item High energy photons emitted in gravitino decays can destroy light elements (D, T, $^{3}$He, $^{4}$He) through photo-dissociation reactions such as those given in Table~\ref{tbl:pdr}; 
\end{enumerate}

\begin{table}[t!]
\tbl{Photo-dissociation reactions that the high energy photons can participate in. The light elements may be destroyed through these reactions and thus their abundance may be changed.}
{\begin{tabular}{@{}clcrr@{}}
\toprule
\hspace{0.15in} & reaction & \hspace{0.25in} & threshold (MeV) & \hspace{0.15in}\\
\colrule
& $D+\gamma \rightarrow n+p$ & & $2.225$ &\\
& $T+\gamma \rightarrow n+D$ & & 6.257 & \\
& $T + \gamma \rightarrow p + n + n$ & & 8.482 &\\
& $^{3} He + \gamma \rightarrow p + D$ & & 5.494 &\\
& $^{4} He + \gamma \rightarrow  p + T$ & & 19.815 &\\
& $^{4} He + \gamma \rightarrow n + ^{3}He$ & & 20.578 &\\
& $^{4} He + \gamma \rightarrow p + n + D$ & & 26.072 &\\
\hline
\end{tabular}}
\label{tbl:pdr}
\end{table}

The gravitino number density, $n_{3/2}$, during the thermalization stage after the inflation is governed by the following Boltzmann equation~\cite{Giudice:1999am}, 
\begin{equation}
\frac{d}{dt} n_{3/2} + 3 H n_{3/2} \simeq \left< \sum_{\mbox{\tiny tot}} v \right> \cdot n_{\mbox{\tiny light}}^{2}
\end{equation}
where $\sum_{\mbox{\tiny tot}} \sim 1/M_{\mbox{\tiny Pl}}^{2}$ is the total cross section determining the production rate of gravitinos
and $n_{\mbox{\tiny light}} \sim T^{3}$ is the number density of light particles in the thermal bath. 
\begin{figure}[t!]
\centerline{\psfig{file=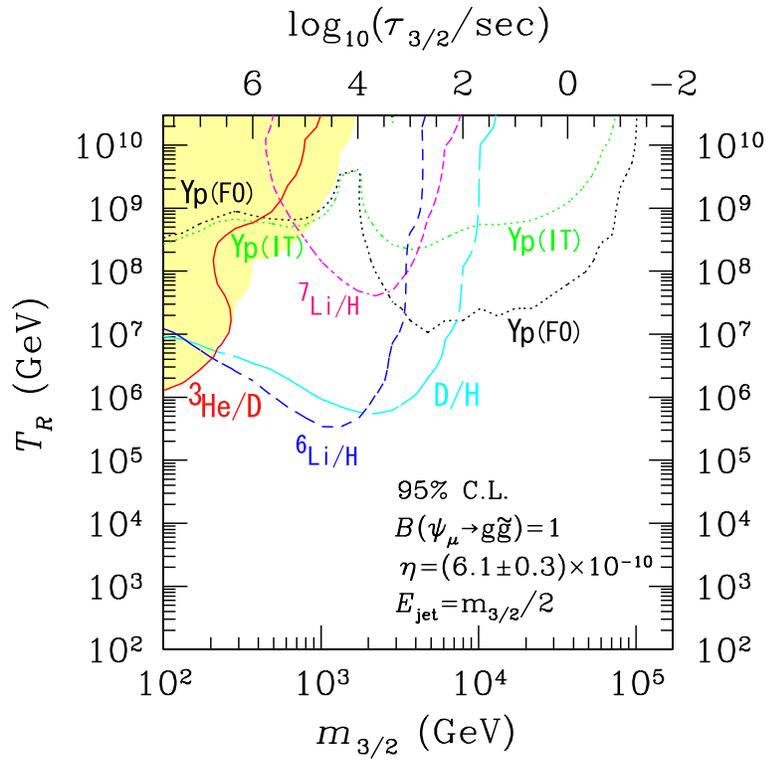,width=10.0cm}}
\caption{Upper bound on reheating temperature as a function of the gravitino mass, for the case when gravitino dominant decays into a gluon-gluino pair. 
Figure taken from Ref.~\cite{Kawasaki:2004qu}.}
\label{fig:gravitino-g}
\end{figure}
As the thermalization is very fast, the friction term $3H n_{3/2}$ in the above Boltzmann equation can be neglected. Using the fact that the Universe is radiation dominant, $H \sim t^{-1} \sim T^{2}/M_{\mbox{\tiny Pl}}$, it follows that,
\begin{equation}
n_{3/2} \sim \frac{T^{4}}{M_{\mbox{\tiny Pl}}} \; ,
\end{equation}
and the number density at thermalization in unit of entropy then reads,
\begin{equation}
\frac{n_{3/2}}{s} \simeq 10^{-2}\frac{T_{RH}}{M_{\mbox{\tiny Pl}}} \; .
\end{equation}
The observed abundances for various light elements are, 
\begin{eqnarray}
0.22 < Y_{p} = (\rho_{^{4}He}/\rho_{B})_{p} < 0.24 \; ,\\
(n_{D}/n_{H}) > 1.8 \times 10^{-5} \; , \\
\biggl(\frac{n_{D}+n_{^{3}He}}{n_{H}} \biggr)_{p} < 10^{-4} \; .
\end{eqnarray}
The most stringent constraint is from the abundance of (D + $^{3}$He) which requires the gravitino number density to be
\begin{equation}
\frac{n_{3/2} }{ s} \simeq 10^{-2} \frac{T_{RH}}{M_{\mbox{\tiny Pl}}} \lesssim 10^{-12} \; .
\end{equation}
The constraint $T_{RH} < 10^{8-9} \; \mbox{GeV}$ then follows. 
More recently, it has been shown that, for hadronic decay modes, 
$\psi \rightarrow g + \tilde{g}$, the bounds are even more stringent, $T_{R} < 10^{6-7}$ GeV, for gravitino mass $m_{3/2} \sim 100$ GeV~\cite{Kawasaki:2004qu}.
Fig.~\ref{fig:gravitino-g} shows the upper bound on the reheating temperature, $T_{R}$, for different values of gravitino mass, $m_{3/2}$. 
Table~\ref{tbl:trg} summarizes the numerical results for the upper bound on $T_{R}$ for various values of $m_{3/2}$. It has also been pointed out~\cite{Pradler:2006hh} that including recent constraint from $^{6}Li$, a new upper bound $T_{R} < 10^{7}$ GeV can be derived for the case of gravitino LSP in the constrained minimal supersymmetric Standard Model (CMSSM).

\begin{table}[t!]
\tbl{Upper bound on the reheating temperature for different values of gravitino mass.}
{\begin{tabular}{@{}crcrc@{}}
\toprule
\hspace{0.15in} & Gravitino mass $m_{3/2}$ & \hspace{0.25in}  & Upper bound on $T_{R}$ & \hspace{0.15in}\\
\colrule
&$\lesssim 100 \; $ GeV  $\qquad$ & & $10^{6-7} \;$ GeV &\\
&$100 \;$ GeV $ -  1 \;$ TeV $\qquad$ & & $10^{7-9} \;$ GeV &\\
&$1 \; $ TeV  $- 3 \;$  TeV $\qquad$ & & $10^{9-12} \; $ GeV &\\
&$3 \; $ TeV $ - 10 \;$ TeV  $\qquad$ & & $10^{12}\;$ GeV &\\
\botrule
\end{tabular}}
\label{tbl:trg}
\end{table}%

There is therefore a conflict between generation of sufficient amount of leptogenesis and not overly producing gravitinos. To avoid these conflicts, various non-standard scenarios for leptogenesis have been proposed.  These are discussed in the next section.

\section{Non-standard Scenarios}\label{sec:alternate}

There are a few non-standard scenarios proposed to evade the gravitino over-production problem. In these new scenarios, the conflicts between leptogenesis and gravitino over-production problem are overcome by,  (i) resonant enhancement in the self-energy diagrams due to near degenerate right-handed neutrino masses (resonant leptogenesis);  (ii) relaxing the relation between the lepton number asymmetry and the right-handed neutrino mass (soft leptogenesis);  (iii) relaxing the relation between the reheating temperature and the right-handed neutrino mass (non-thermal leptogenesis). These scenarios are discussed below.

\subsection{Resonant Leptogenesis}\label{sec:res}

Recall that in the standard leptogenesis discussed in Sec.~\ref{sec:stlpg}, contributions to the CP asymmetry is due to the interference between the tree-level and the one-loop diagrams, that include the vertex correction and self-energy diagrams. It was pointed out in 
Ref.~\cite{Pilaftsis:1997jf} that in the limit $M_{N_{i}} - M_{N_{j}} \ll M_{N_{i}}$, the self-energy diagrams dominate,
\begin{equation}
\epsilon_{N_{i}}^{\mbox{\tiny Self}} = \frac{Im[ (h_{\nu} h_{\nu}^{\dagger})_{ij} ]^{2}}{(h_{\nu}h_{\nu}^{\dagger})_{ii} 
(h_{\nu}h_{\nu}^{\dagger})_{jj}} 
\biggl[ \frac{(M_{i}^{2}-M_{j}^{2})M_{i} \Gamma_{N_{j}}}{(M_{i}^{2}-M_{j}^{2})^{2} + M_{i}^{2} 
\Gamma_{N_{j}}^{2}} \biggr] \; .
\end{equation}
When the lightest two RH neutrinos have near degenerate masses, $M_{1}^{2} - M_{2}^{2} \sim \Gamma_{N_{2}}^{2}$, the asymmetry can be enhanced. To be more specific, CP asymmetry of $\mathcal{O}(1)$ is possible, when
\begin{equation}
M_{1}-M_{2} \sim \frac{1}{2} \Gamma_{N_{1,2}} \quad , \quad \mbox{assuming~~}
\frac{Im(h_{\nu}h_{\nu}^{\dagger})_{12}^{2}}{(h_{\nu}h_{\nu}^{\dagger})_{11} (h_{\nu} h_{\nu}^{\dagger})_{22}} \sim 1\; .
\end{equation}
Due to this resonant effect, the bound on the RH neutrino mass scale from the requirement of generating sufficient lepton number asymmetry can be significantly lower. It has been shown that sufficient baryogenesis can be obtained even with $M_{1,2} \sim $ TeV~\cite{Pilaftsis:2003gt}.

\subsection{Soft Leptogenesis}\label{sec:softlept}

CP violation in leptogenesis can arise in two ways: it can arise in decays, which is the case in standard leptogenesis described in the previous section. It can also arise in mixing. An example of this is the soft leptogenesis. Recall that in the Kaon system, non-vanishing CP violation exists due to the mismatch between CP eigenstates and mass eigenstates (for a review, see for example, Ref.~\cite{Nir:2001ge}). 
The CP eigenstates of the $K^{0}$ system are $\frac{1}{\sqrt{2}} \bigl( \big| K^{0} \big> \pm \big| \overline{K}^{0} \big> \bigr)$. 
The time evolution of the $(K^{0},\overline{K}^{0})$ system is described by the following Schr\"odinger equation, 
\begin{equation} 
\frac{d}{dt} \left( \begin{array}{c}
K^{0} \\ \overline{K}^{0}
\end{array}\right)
=  \mathcal{H} \left( \begin{array}{c}
K^{0} \\ \overline{K}^{0}
\end{array}\right)
\end{equation}
where the Hamiltanian $\mathcal{H}$ is given by $\mathcal{H}=\mathcal{M}-\frac{i}{2} \mathcal{A}$. 
Here, the off-diagonal matrix element  $\mathcal{M}_{12}$ describes the dispersive part of the transition amplitude, while the element 
$\mathcal{A}_{12}$ gives the absorptive part of the amplitude. The physical (mass) eigenstates, $\big|K_{L,S}\big>$, are given in terms of the flavor eigenstates, $\big| K^{0} \big>$ and $\big| \overline{K}^{0}\big>$, as
\begin{eqnarray}
\big| K_{L} \big>  & = & p \big| K^{0} \big> + q \big| \overline{K}^{0} \big>
\\
\big| K_{S} \big>  & = & p \big| K^{0} \big> - q \big| \overline{K}^{0} \big> \; .
\end{eqnarray}
To have non-vanishing CP violation requires that there exists a mismatch between the CP eigenstates and the physical eigenstates. This in turn implies, 
\begin{equation}
\bigg| \frac{q}{p} \bigg| \ne 1 \; ,  \quad \mbox{where} \quad  
\biggl(\frac{q}{p}\biggr)^{2} = \biggl( \frac{2\mathcal{M}_{12}^{\ast} - i \mathcal{A}_{12}^{\ast}}{2\mathcal{M}_{12} - i \mathcal{A}_{12}}\biggr) \; .
\end{equation}

For soft leptogenesis, the relevant soft SUSY Lagrangian  
that involves lightest RH sneutrinos $\widetilde{\nu}_{R_{1}}$ 
is the following,
\begin{eqnarray}
-\mathcal{L}_{soft}  &=&  \biggl(\frac{1}{2} B M_{1} \widetilde{\nu}_{R_{1}} 
\widetilde{\nu}_{R_{1}} 
+ A \mathcal{Y}_{1i} \widetilde{L}_{i} \widetilde{\nu}_{R_{1}} H_{u} + h.c.\biggr) 
\nonumber\\
&&
+ \widetilde{m}^{2} \widetilde{\nu}_{R_{1}}^{\dagger} 
\widetilde{\nu}_{R_{1}} \; .
\end{eqnarray}
This soft SUSY Lagrangian and the superpotential that involves 
the lightest RH neutrino, $N_{1}$, 
\begin{equation}
W = M_{1} N_{1} N_{1} + \mathcal{Y}_{1i} L_{i} N_{1} H_{u}
\end{equation}
give rise to the following interactions 
\begin{eqnarray}
-\mathcal{L}_{\mathcal{A}}  =  
\widetilde{\nu}_{R_{1}} (
M_{1} Y_{1i}^{\ast} \widetilde{\ell}_{i}^{\ast} H_{u}^{\ast}
+\mathcal{Y}_{1i} \overline{\widetilde{H}}_{u} \ell_{L}^{i} 
+ A \mathcal{Y}_{1i} \widetilde{\ell}_{i} H_{u}
) + h. c. \quad ,
\end{eqnarray}
and mass terms (to leading order in soft SUSY breaking terms),
\begin{equation}
-\mathcal{L}_{\mathcal{M}}  =  
(M_{1}^{2} \widetilde{\nu}_{R_{1}}^{\dagger} \widetilde{\nu}_{R_{1}}  +  
\frac{1}{2} B M_{1} 
\widetilde{\nu}_{R_{1}} \widetilde{\nu}_{R_{1}} ) + h.c. \; .
\end{equation}
Diagonalization of the mass matrix $\mathcal{M}$ for  
the two states $\widetilde{\nu}_{R_{1}}$ and 
$\widetilde{\nu}_{R_{1}}^{\dagger}$ 
leads to eigenstates  
$\widetilde{N}_{+}$ and $\widetilde{N}_{-}$ 
with masses,  
\begin{equation}
M_{\pm} \simeq M_{1} \biggl(1 \pm \frac{|B|}{2M_{1}} \biggr) \; ,
\end{equation}
where the leading order term $M_{1}$ is the F-term contribution 
from the superpotential (RH neutrino mass term) and    
the mass difference between the two mass eigenstates $\widetilde{N}_{+}$ 
and $\widetilde{N}_{-}$ is induced by the SUSY breaking $B$ term.
The time evolution of the 
$\widetilde{\nu}_{R_{1}}$-$\widetilde{\nu}^{\dagger}_{R_{1}}$ system  
is governed by the Schr\"{o}dinger equation, 
\begin{equation}
\frac{d}{dt} \left(
\begin{array}{c}
\widetilde{\nu}_{R_{1}}\\
\widetilde{\nu}_{R_{1}}^{\dagger}
\end{array}\right)
= \mathcal{H}
\left(
\begin{array}{c}
\widetilde{\nu}_{R_{1}}\\
\widetilde{\nu}_{R_{1}}^{\dagger}
\end{array}\right) \; ,
\end{equation}
where the Hamiltonian  is $\mathcal{H}= \mathcal{M} - \frac{i}{2} \mathcal{A}$ with $\mathcal{M}$ and $\mathcal{A}$ 
being~\cite{Grossman:2003jv,D'Ambrosio:2003wy}, 
\begin{eqnarray}
\mathcal{M} & = & \left(
\begin{array}{cc}
1 & \frac{B^{\ast}}{2M_{1}}\\
\frac{B}{2M_{1}} & 1
\end{array}\right) \; M_{1} \; ,
\\
\mathcal{A} & = & \left(
\begin{array}{cc}
1 & \frac{A^{\ast}}{M_{1}}\\
\frac{A}{M_{1}} & 1
\end{array}\right) \Gamma_{1} \; .
\end{eqnarray}
For the decay of the lightest RH sneutrino, $\widetilde{\nu}_{R_{1}}$, 
the total decay width  $\Gamma_{1}$ is given by, in the basis where both the charged lepton 
mass matrix and the RH neutrino mass matrix are diagonal,
\begin{equation}
\Gamma_{1} =  \frac{1}{4\pi} 
(\mathcal{Y}_{\nu}\mathcal{Y}_{\nu}^{\dagger})_{11} M_{1}
\; .
\end{equation}
The eigenstates of the Hamiltonian $\mathcal{H}$ are  
$\widetilde{N}_{\pm}^{\prime} = p \widetilde{N} 
\pm q \widetilde{N}^{\dagger}$, where $|p|^{2} + |q|^{2} = 1$.
The ratio $q/p$ is given in terms of $\mathcal{M}$ and $\Gamma_{1}$ as,
\begin{eqnarray}
\biggl( \frac{q}{p} \biggr)^{2} =  
\frac{2\mathcal{M}_{12}^{\ast} - i \mathcal{A}_{12}^{\ast}}
{2\mathcal{M}_{12} - i \mathcal{A}_{12}} \simeq 1 + \mbox{Im} \biggl( \frac{2\Gamma_{1} A}{BM_{1}} \biggr) \; ,
\end{eqnarray}
in the limit $\mathcal{A}_{12} \ll \mathcal{M}_{12}$.
Similar to the $K^{0}-\overline{K}^{0}$ system, 
the source of CP violation in the lepton number asymmetry 
considered here is due to 
the CP violation in the mixing which occurs when the two neutral 
mass eigenstates ($\widetilde{N}_{+}$, $\widetilde{N}_{-}$), 
are different from the interaction eigenstates,
($\widetilde{N}^{\prime}_{+}$, $\widetilde{N}^{\prime}_{-}$).
Therefore CP violation in mixing is present
as long as the quantity $|q/p| \ne 1$, which requires 
\begin{equation}
\mbox{Im} \biggl( \frac{A\Gamma_{1}}{M_{1} B} \biggr) \ne 0 \; . 
\end{equation} 
For this to occur, SUSY breaking, {\it i.e.} non-vanishing $A$ 
{\it and} $B$, is required.
As the relative phase between the parameters $A$ and $B$ 
can be rotated away by an $U(1)_{R}$-rotation as discussed in Sec.~\ref{sec:counting}, without loss of generality  
we assume from now on that the remaining physical phase  
is solely coming from the tri-linear coupling, $A$. 

The total lepton number asymmetry integrated over time, $\epsilon$,  
is defined as the ratio of the difference to the sum of the decay widths $\Gamma$ 
for $\widetilde{\nu}_{R_{1}}$ and $\widetilde{\nu}_{R_{1}}^{\dagger}$ 
into final states of the slepton doublet $\widetilde{L}$ and the Higgs doublet $H$, 
or the lepton doublet $L$ and the Higgsino $\widetilde{H}$ or their conjugates,
\begin{equation}
\epsilon = \frac{\sum_{f} \int_{0}^{\infty} [
\Gamma(\widetilde{\nu}_{R_{1}}, \widetilde{\nu}_{R_{1}}^{\dagger} \rightarrow
f) - 
\Gamma(\widetilde{\nu}_{R_{1}}, \widetilde{\nu}_{R_{1}}^{\dagger} \rightarrow
\overline{f})]}
{\sum_{f} \int_{0}^{\infty} 
[\Gamma( \widetilde{\nu}_{R_{1}}, \widetilde{\nu}_{R_{1}}^{\dagger} 
\rightarrow f) + 
\Gamma(\widetilde{\nu}_{R_{1}}, \widetilde{\nu}_{R_{1}}^{\dagger}
\rightarrow \overline{f})]} \; .
\end{equation}
Here the final states $f = (\widetilde{L}\; H), 
\; (L \; \widetilde{H})$ have lepton number
$+1$, and $\overline{f}$ denotes their conjugate, $(\widetilde{L}^{\dagger} 
\; H^{\dagger}), 
\; (\overline{L} \; \overline{\widetilde{H}})$, which have lepton number
$-1$.
After carrying out the time integration, the total CP asymmetry 
is~\cite{Grossman:2003jv,D'Ambrosio:2003wy}, 
\begin{equation}
\epsilon = \biggl(
\frac{4\Gamma_{1} B}{\Gamma_{1}^{2}+4B^{2}} \biggr)
\frac{\mbox{Im}(A)}{M_{1}} \delta_{B-F}
\end{equation}
where the additional factor $\delta_{B-F}$ takes into account the thermal effects 
due to the difference between the occupation numbers of bosons and 
fermions~\cite{Covi:1997dr}.

The final result for the baryon asymmetry is~\cite{Grossman:2003jv,D'Ambrosio:2003wy},
\begin{eqnarray}
\frac{n_{B}}{s} & \simeq & 
- c_{s} \; d_{\widetilde{\nu}_{R}} \; 
\epsilon \; \kappa \; ,
\nonumber\\
& \simeq & 
-1.48 \times 10^{-3} \epsilon \; \kappa \; ,
\nonumber\\
& \simeq & -(1.48 \times 10^{-3}) 
\biggl( \frac{\mbox{Im}(A)}{M_{1}} \biggl)
\; R \; \delta_{B-F} \; \kappa \; ,
\end{eqnarray}
where $d_{\widetilde{\nu}_{R}}$ in the first line is the density of the lightest sneutrino 
in equilibrium in units of entropy density, and is given by, $d_{\widetilde{\nu}_{R}} 
= 45 \zeta(3)/(\pi^{4}g_{\ast})$; the factor $c_{s}$, 
which characterizes the amount of $B-L$ asymmetry being converted into the baryon asymmetry $Y_{B}$,  is defined in Eq.~\ref{eq:convert}. 
The parameter $\kappa$ is the efficiency factor given in Sec.~\ref{sec:boltz}. 
The resonance factor $R$ is defined as the following ratio, 
\begin{equation}
R \equiv \frac{4 \Gamma_{1} B}{\Gamma_{1}^{2} + 4 B^{2}} \; ,
\end{equation}
which gives a value equal to one when the resonance condition, $\Gamma_{1} = 2|B|$, 
is satisfied,  
leading to maximal CP asymmetry. As $\Gamma_{1}$ is of the order of 
$\mathcal{O}(0.1-1)$  GeV, to satisfy the resonance condition, 
a small value for $B \ll \widetilde{m}$ is thus needed. 
Such a small value of $B$ can be generated by some dynamical 
relaxation mechanisms~\cite{Yamaguchi:2002zy} 
in which $B$ vanishes in the leading order. A small 
value of $B \sim \widetilde{m}^{2}/M_{1}$ is then generated by an operator 
$\int d^{4} \theta ZZ^{\dagger}N_{1}^{2} / M_{pl}^{2}$ in the K\"{a}hler 
potential, where $Z$ is the SUSY breaking spurion field, 
$Z = \theta^{2}~\widetilde{m} M_{pl}$~\cite{D'Ambrosio:2003wy}.
In a specific SO(10) model constructed in Ref.~\cite{Chen:2000fp,Chen:2004xy}, it has been shown that with the parameter $B^\prime  \equiv \sqrt{BM_{1}}$ having the size of the natural 
SUSY breaking scale $\sqrt{\widetilde{m}^{2}} \sim \mathcal{O}(1)$ TeV, a small 
value for $B$ required by the resonance condition 
$B \sim \Gamma_{1} \sim \mathcal{O}(0.1)$ GeV can be obtained.

\subsection{Non-thermal Leptogenesis}

The conflict between generating sufficient leptogenesis and not overly producing gravitinos in thermal leptogenesis arises due to strong dependence of the reheating temperature $T_{R}$ on the lightest RH mass, $M_{R_{1}}$, in thermal leptogenesis.  This problem may be avoided if the relation between the reheating temperature and the lightest RH neutrino mass is loosened. This is the case if the primordial RH neutrinos are produced non-thermally. One possible way to have non-thermal leptogenesis is to generate the primordial right-handed neutrinos through the inflaton decay~\cite{Fujii:2002jw}. 

Inflation solves the horizon and flatness problem, and it accounts for the origin of density fluctuations. 
Assume that the inflaton decays dominantly into a pair of lightest RH neutrinos, $\Phi \rightarrow N_{1} + N_{1}$. For this decay to occur,  the inflaton mass $m_{\Phi}$ has to be greater than $2 M_{1}$. For simplicity, let us also assume that the decay modes into $N_{2,3}$ are energetically forbidden. The produced $N_{1}$ in inflaton decay then subsequently decays into $H + \ell_{L}$ and $H^{\dagger} + \ell_{L}^{\dagger}$. 
The out-of-equilibrium condition is automatically satisfied, if $T_{R} < M_{1}$. 
The CP asymmetry is generated by the interference of tree level and one-loop diagrams, 
\begin{equation}
\epsilon = -\frac{3}{8\pi} \frac{M_{1}}{\left< H \right>^{2}} m_{3} \delta_{eff} \; ,
\end{equation}
where $\delta_{eff}$ is given in terms of the neutrino Yukawa matrix elements and light neutrino masses as,
\begin{equation}
\delta_{eff} = \frac{Im\bigg\{ h_{13}^{2} + \frac{m_{2}}{m_{3}} h_{12}^{2} + \frac{m_{1}}{m_{3}} h_{11}^{2}\bigg\}}{\big|h_{13}\big|^{2} + \big|h_{12}\big|^{2} + \big| h_{11}\big|^{2}} \; .
\end{equation}  
Numerically, the asymmetry is given by~\cite{Fujii:2002jw},
\begin{equation}
\epsilon \simeq -2 \times 10^{-6} \biggl( \frac{M_{1}}{10^{10} \; \mbox{GeV}} \biggr) \biggl( \frac{m_{3}}{0.05 \; \mbox{eV}}\biggr) \delta_{eff} \; .
\end{equation}

The chain decays $\Phi \rightarrow N_{1} + N_{1}$ and $N_{1} \rightarrow H + \ell_{L}$ or $H^{\dagger} + \ell_{L}^{\dagger}$ reheat the Universe producing not only the lepton number asymmetry but also the entropy for the thermal bath. Taking such effects into account, 
the ratio of lepton number to entropy density after the reheating~\cite{Fujii:2002jw} is then,
\begin{equation}
\frac{n_{L}}{s} \simeq -\frac{3}{2} \epsilon \frac{T_{R}}{m_{\Phi}} \simeq 3 \times 10^{-10} \biggl( 
\frac{T_{R}}{10^{6} \; \mbox{GeV}} \biggr) \biggl( \frac{M_{1}}{m_{\Phi}}\biggr) \biggl( \frac{m_{3}}{0.05 \; \mbox{eV}}\biggr) \; ,
\end{equation}
assuming $\delta_{eff}=1$. The ratio $n_{B}/s \sim 10^{-10}$ can thus be obtained with $M_{1} \lesssim m_{\Phi}$, and $T_{R} \lesssim 10^{6}$ GeV.

\section{Connection between leptogenesis and neutrino oscillation}\label{sec:connect}

As mentioned in Sec.~\ref{sec:counting}, there is generally no connection between low energy CP violating processes, such as CP violation in neutrino oscillation and   in neutrinoless double beta decay, and leptogenesis, which occurs at very high energy scale. This is due to the extra phases and mixing angles present in the heavy neutrino sector. One way to establish such connection is by reducing the inter-family couplings (equivalently, by imposing texture zero in the Yukawa matrix). This is the case for the $3\times 2$ seesaw model. A more powerful way to obtain such connection is to have all CP violation, both low energy and high energy, come from the same origin. This ensues if CP violation occurs spontaneously.    
Below we described these two models in which such connection does exist. 

\subsection{Models with Two Right-Handed Neutrinos}

One type of models where there exists connection between CP violating processes at high and low energies is models with only two RH neutrinos. 
In this case, the neutrino Dirac mass matrix is a $3 \times 2$  matrix. This $3\times 2$ Yukawa matrix has six complex parameters, and hence six phases, out of which, three can be absorbed by the wave functions of the three charged leptons. Even though, the reduction in the number of right-handed neutrinos reduces the number of CP phases in high energy, it also reduces the number of CP phases at low energy to two. There is therefor still one high energy phase that cannot be determined by measuring the low energy phases. However, if one further assumes that the $3\time 2$  Yukawa matrix has two zeros, there is then only one CP phase in the Yukawa matrix, making the existence of the connection possible. 

The existence of two right-handed neutrinos is required 
by the cancellation of Witten anomaly, if a global leptonic $SU(2)$ family symmetry is 
imposed~\cite{Kuchimanchi:2002yu}. (For implications of non-anomalous gauge symmetry for neutrino masses, see Ref.~\cite{Chen:2006hn}. This model provided the interesting possibility of probing the neutrino sector at the colliders through their couplings to the $Z^{\prime}$ gauge 
boson~\cite{Rizzo:2006nw}.) 
Along this line, Frampton, Glashow and Yanagida proposed a model, which has 
the following Lagrangian~\cite{Frampton:2002qc}, 
\begin{equation}\label{FGY}
\mathcal{L} = \frac{1}{2} (N_{1} N_{2}) 
\left(\begin{array}{ccc}
M_{1} & 0\\
0 & M_{2}
\end{array}\right)
\left(\begin{array}{c} N_{1} \\ N_{2} \end{array}\right)
+ (N_{1} N_{2}) \left(\begin{array}{ccc}
a & a^{\prime} & 0\\
0 & b & b^{\prime}
\end{array}\right)
\left(\begin{array}{c} \ell_{1} \\ \ell_{2} \\ \ell_{3}
\end{array}\right) H + h.c. \; ,
\end{equation} 
with the Yukawa matrix having two zeros in the $N_{1}-\ell_{3}$ and $N_{2}-\ell_{1}$ couplings. 
The effective neutrino mass matrix due to this Lagrangian is obtained, 
using the see-saw formula,
\begin{equation}
\left(\begin{array}{ccc}
\frac{a^{2}}{M_{1}} & \frac{aa^{\prime}}{M_{1}} & 0
\\
\frac{aa^{\prime}}{M_{1}} & \frac{a^{\prime 2}}{M_{1}} + \frac{b^{2}}{M_{2}} & \frac{bb^{\prime}}{M_{2}}
\\
0 & \frac{bb^{\prime}}{M_{2}} & \frac{b^{\prime2}}{M_{2}}
\end{array}\right),
\end{equation}
where $a, b, b^{\prime}$ are real and $a^{\prime} = |a^{\prime}| e^{i\delta}$. 
By takinging all of them to be real, with the choice $a^{\prime} = \sqrt{2} a$ 
and $b=b^{\prime}$, and assuming $a^{2}/M_{1} \ll b^{2}/M_{2}$, 
the effective neutrino masses and mixing matrix are obtained
\begin{equation}
m_{\nu_{1}} = 0, \quad m_{\nu_{2}} = \frac{2a^{2}}{M_{1}}, \quad 
m_{\nu_{3}} = \frac{2b^{2}}{M_{2}}
\end{equation}
\begin{equation}
U = \left(\begin{array}{ccc}
1/\sqrt{2} & 1/\sqrt{2} & 0\\
-1/2 & 1/2 & 1/\sqrt{2}\\
1/2 & -1/2 & 1/\sqrt{2}
\end{array}\right) \times
\left(\begin{array}{ccc}
1 & 0 & 0\\
0 & \cos\theta & \sin\theta \\
0 & -\sin\theta & \cos\theta
\end{array}\right),
\end{equation}
where $\theta \simeq m_{\nu_{2}}/\sqrt{2}m_{\nu_{3}}$, and the 
observed bi-large mixing angles and 
$\Delta m_{atm}^{2}$ and $\Delta m_{\odot}^{2}$ can be accommodated. 
An interesting feature of this model 
is that the sign of the baryon number asymmetry ($B \propto \xi_{B} = 
Y^{2} a^{2} b^{2} \sin2\delta$) 
is related to the sign of the 
CP violation in neutrino oscillation ($\xi_{osc}$) in the following way 
\begin{equation}
\xi_{osc} = -\frac{a^{4}b^{4}}{M_{1}^{3}M_{2}^{3}} (2 + Y^{2}) \xi_{B} \propto - B
\end{equation}
assuming the baryon number asymmetry is resulting from leptogenesis 
due to the decay of the 
lighter one of the two heavy neutrinos, $N_{1}$. This idea can be  realized in a SO(10) with additional singlets~\cite{Raby:2003ay}.

\subsection{Models with Spontaneous CP Violation (\& Triplet Leptogenesis)}

The second type of models in which relation between leptogenesis and low energy CP violation exists is the minimal left-right symmetric model with spontaneous CP violation (SCPV)~\cite{Chen:2004ww}. 
The left-right (LR) model~\cite{Pati:1974yy}  is based on the gauge group, $SU(3)_{c} \times SU(2)_{L} \times SU(2)_{R} \times U(1)_{B-L} \times P$, where the parity $P$ acts on the two $SU(2)$'s. (See also Kaladi Babu's lectures.) In this model, the electric charge $Q$ can be understood as the sum of the two $T^{3}$ quantum numbers of the $SU(2)$ gauge groups,
\begin{equation}
Q = T_{3,L} + T_{3,R} + \frac{1}{2} (B-L) \; .
\end{equation}
The {\it minimal} LR model has the following particle content: In the fermion sector, the iso-singlet quarks form a doublet under $SU(2)_{R}$, and similarly for $e_{R}$ and $\nu_{R}$,
\begin{eqnarray}
Q_{i,L} & = \left(\begin{array}{c}
u \\ d
\end{array}\right)_{i,L} \sim (1/2,0,1/3), \qquad 
Q_{i,R} & = \left(\begin{array}{c}
u \\ d
\end{array}\right)_{i,R} \sim (0,1/2,1/3)\nonumber\\
L_{i,L} & = \left(\begin{array}{c}
e \\ \nu
\end{array}\right)_{i,L} \sim (1/2,0,-1), \qquad
L_{i,R} & = \left(\begin{array}{c}
e \\ \nu
\end{array}\right)_{i,R} \sim (0,1/2,-1) \; .
\nonumber
\end{eqnarray}
In the scalar sector, there is a bi-doublet and one triplet for each of the $SU(2)$'s,
\begin{eqnarray}
\Phi & = & \left(
\begin{array}{cc}
\phi^{0}_{1}  & \phi_{2}^{+}\\
\phi_{1}^{-} & \phi_{2}^{0}
\end{array}\right)
\sim (1/2, \; 1/2, \; 0)
\nonumber\\
\Delta_{L} & = & \left(
\begin{array}{cc}
\Delta^{+}_{L} /\sqrt{2} & \Delta_{L}^{++}\\
\Delta_{L}^{0} & - \Delta_{L}^{+}/\sqrt{2}
\end{array}\right)
\sim (1, \; 0, \; 2)\nonumber\\ 
\Delta_{R}  & = &  \left(
\begin{array}{cc}
\Delta^{+}_{R} /\sqrt{2} & \Delta_{R}^{++}\\
\Delta_{R}^{0} & - \Delta_{R}^{+}/\sqrt{2}
\end{array}\right)
\sim (0, \; 1, \; 2) \nonumber \; .
\end{eqnarray}
Under the parity $P$, these fields transform as, 
\begin{equation}
\Psi_{L} \leftrightarrow \Psi_{R}, \quad 
\Delta_{L} \leftrightarrow \Delta_{R}, \quad
\Phi \leftrightarrow \Phi^{\dagger} \; .
\end{equation}
The VEV of the $SU(2)_{R}$ breaks the left-right symmetry down to the SM gauge group,
\begin{eqnarray}
SU(3)_{c} \times SU(2)_{L} \times SU(2)_{R} \times U(1)_{B-L} \times P
\nonumber\\
\rightarrow SU(3)_{c} \times SU(2)_{L} \times U(1)_{Y} \; ,
\end{eqnarray}
and the subsequent breaking of the electroweak symmetry is achieved by the bi-doublet VEV. 
In general, 
\begin{eqnarray}
\left<\Phi\right> & = & \left(\begin{array}{cc}
\kappa e^{i\alpha_{\kappa}} & 0 \\
0 & \kappa^{\prime} e^{i\alpha_{\kappa^{\prime}}}\end{array}\right),
\;  \\
\left< \Delta_{L} \right> & = & 
\left(\begin{array}{cc}
0 & 0 \\
v_{L} e^{i\alpha_{L}} & 0
\end{array}\right), \; 
\left< \Delta_{R}\right> = 
\left(\begin{array}{cc}
0 & 0 \\
v_{R} e^{i\alpha_{R}} & 0
\end{array}\right) \; . \nonumber
\end{eqnarray}
To get realistic SM gauge boson masses, the VEV's of the bi-doublet Higgs 
must satisfy 
$v^{2} \equiv 
|\kappa|^2 + |\kappa^{'}|^{2} \simeq 2 M_{w}^{2}/g^{2} \simeq (174 \mbox{GeV})^{2}$.
Generally, a non-vanishing VEV for the 
$SU(2)_{L}$ triplet Higgs is induced, and it is suppressed 
by the heavy $SU(2)_{R}$ breaking scale similar 
to the see-saw mechanism for the neutrinos,  
\begin{equation}
<\Delta_{L}> =  \left(
\begin{array}{cc}
0 & 0 \\
v_{L} e^{i\alpha_{L}} & 0
\end{array}\right) \; , \qquad v_{L}v_{R} 
= \beta |\kappa|^{2}  \; ,
\end{equation}
where the parameter $\beta$ is a function of the order 
$\mathcal{O}(1)$ coupling constants in the scalar potential 
and $v_{R}$, $v_{L}$, $\kappa$ and $\kappa^{\prime}$ 
are positive real numbers in the above equations. (The presence of a triplet Higgs in warped extra dimensions can provide a natural 
way to generate small Majorana masses for the neutrinos~\cite{Chen:2005mz}.)  
Due to this see-saw suppression, for a $SU(2)_{R}$ breaking scale as high as 
$10^{15}$ GeV, which is required by the smallness of the neutrino masses,  
the induced $SU(2)_{L}$ triplet VEV is well below the upper bound 
set by the electroweak precision 
constraints~\cite{Blank:1997qa}.  
The scalar potential that gives rise to the 
vacuum alignment described can be found in 
Ref.~\cite{Deshpande:1990ip}.

The Yukawa sector of the model is given by
$\mathcal{L}_{Yuk} = \mathcal{L}_{q} + \mathcal{L}_{\ell}$, 
where $\mathcal{L}_{q}$ and $\mathcal{L}_{\ell}$ are 
the Yukawa interactions in the quark and lepton sectors, 
respectively. The Lagrangian for quark Yukawa interactions is 
given by,  
\begin{equation}
-\mathcal{L}_{q} = \overline{Q}_{i,R} (F_{ij} \Phi + G_{ij} 
\tilde{\Phi}) Q_{j,L} + h.c.
\end{equation}
where $\tilde{\Phi} \equiv \tau_{2} \Phi^{\ast} \tau_{2}$. 
In general, $F_{ij}$ and $G_{ij}$ 
are Hermitian to preseve left-right symmetry. Because of our assumption  of 
SCPV with complex vacuum expectation values, 
the matrices $F_{ij}$ and $G_{ij}$ are real. 
The Yukawa interactions responsible for generating the lepton masses 
are summarized in the following  Lagrangian, 
$\mathcal{L}_{\ell}$,
\begin{eqnarray}
-\mathcal{L}_{\ell} & = &  
\overline{L}_{i,R} (P_{ij} \Phi + R_{ij} \tilde{\Phi}) L_{j,L} 
\\
&& + i f_{ij} (L_{i,L}^{T} C\tau_{2} \Delta_{L} L_{j,L} 
+ L_{i,R}^{T} C\tau_{2} \Delta_{R} L_{j,R}) 
+ h.c. \; , \nonumber
\end{eqnarray}
where $\mathcal{C}$ is the Dirac charge conjugation operator, and the matrices 
$P_{ij}$, $R_{ij}$ and $f_{ij}$ are real due to the assumption of SCPV.
Note that the Majorana mass terms $L_{i,L}^{T} \Delta_{L} L_{j,L}$ and 
$L_{i,R}^{T} \Delta_{R} L_{j,R}$ have identical coupling because 
the Lagrangian must be invariant under interchanging 
$L \leftrightarrow R$.
The complete Lagrangian of the model is invariant under 
the unitary transformation, 
under which the matter fields transform as
\begin{equation}
\psi_{L} \rightarrow U_{L} \psi_{L}, \qquad
\psi_{R} \rightarrow U_{R} \psi_{R}
\end{equation}
where $\psi_{L,R}$ are left-handed (right-handed) fermions, 
and the scalar fields transform according to
\begin{equation}
\Phi \rightarrow U_{R} \Phi U_{L}^{\dagger}
, \qquad 
\Delta_{L} \rightarrow U_{L}^{\ast} \Delta_{L} U_{L}^{\dagger}
, \qquad 
\Delta_{R} \rightarrow U_{R}^{\ast} \Delta_{R} U_{R}^{\dagger}
\end{equation}
with the unitary transformations $U_{L}$ and $U_{R}$ being 
\begin{equation}\label{unit}
U_{L}  =  
\left(
\begin{array}{cc}
e^{i\gamma_{L}} & 0\\
0 & e^{-i\gamma_{L}}
\end{array}
\right)
, \qquad 
U_{R} = 
\left(
\begin{array}{cc}
e^{i\gamma_{R}} & 0\\
0 & e^{-i\gamma_{R}}
\end{array}
\right) \; .
\end{equation}
Under these unitary transformations, the VEV's transform as
\begin{eqnarray}
\kappa  \rightarrow  \kappa e^{-i(\gamma_{L}-\gamma_{R})}
, \quad 
\kappa^{\prime} \rightarrow  \kappa^{\prime} e^{i(\gamma_{L}-\gamma_{R})}
, \\
v_{L}  \rightarrow  v_{L} e^{-2i\gamma_{L}}
, \quad 
v_{R} \rightarrow  v_{R} e^{-2i\gamma_{R}} \; . \nonumber
\end{eqnarray}
Thus by re-defining the phases of matter fields with the choice of  
$\gamma_{R}  =  \alpha_{R}/2$ and
$\gamma_{L} =   \alpha_{\kappa} + \alpha_{R}/2$ 
in the unitary matrices $U_{L}$ and $U_{R}$, 
we can rotate away two of the complex phases in the VEV's of 
the scalar fields and are left with only two genuine CP violating phases, 
$\alpha_{\kappa^\prime}$ and $\alpha_{L}$, 
\begin{eqnarray}
<\Phi>  & = & \left(
\begin{array}{cc}
\kappa & 0\\
0 & \kappa^{\prime}e^{i\alpha_{\kappa^{\prime}}}
\end{array}
\right), \quad \\
<\Delta_{L}> & = &  
\left(
\begin{array}{cc}
0 & 0 \\
v_{L}e^{i\alpha_{L}} & 0
\end{array}\right), \quad
<\Delta_{R}> = 
\left(
\begin{array}{cc}
0 & 0 \\
v_{R} & 0
\end{array}\right). \nonumber
 \end{eqnarray}

The quark Yukawa interaction $\mathcal{L}_{q}$ gives rise 
to quark masses after the bi-doublet acquires VEV's
\begin{equation}
M_{u} = \kappa F_{ij} + \kappa^{\prime}  
e^{-i \alpha_{\kappa^\prime}} G_{ij}, 
\quad 
M_{d} = \kappa^{\prime} e^{i\alpha_{\kappa^\prime}}  F_{ij} 
+ \kappa G_{ij} \; .
\end{equation}
Thus the relative phase in the two VEV's in the SU(2) 
bi-doublet, $\alpha_{\kappa^\prime}$, gives rise 
to the CP violating phase in the CKM matrix. 
To obtain realistic quark masses and CKM matrix elements, 
it has been shown that the VEV's of the bi-doublet 
have to satisfy $\kappa/\kappa^\prime \simeq m_{t}/m_{b} \gg 1$~\cite{Ball:1999mb}.
When the triplets and the bi-doublet acquire VEV's, we obtain the following 
mass terms for the leptons
\begin{eqnarray}
M_{e} = \kappa^{\prime} e^{i\alpha_{\kappa^\prime}} P_{ij} + \kappa R_{ij}, 
& \quad
M_{\nu}^{Dirac} = \kappa P_{ij} 
+ \kappa^{\prime} e^{-i\alpha_{\kappa^\prime}} R_{ij} \\
M_{\nu}^{RR} = v_{R} f_{ij}, & \quad
M_{\nu}^{LL} = v_{L} e^{i\alpha_{L}} f_{ij} \; .
\end{eqnarray}
The effective neutrino mass matrix, $M_{\nu}^{\mbox{\tiny eff}}$, which 
arises from the Type-II seesaw mechanism,  
is thus given by
\begin{eqnarray}
M_{\nu}^{eff} & = & M_{\nu}^{II} - M_{\nu}^{I} = (fe^{i\alpha_{L}} - \frac{1}{\beta} P^{T} f^{-1} P)v_{L} \; ,
\\
M_{\nu}^{I} & = & (M_{\nu}^{Dirac})^{T} (M_{\nu}^{RR})^{-1} (M_{\nu}^{Dirac})\\
& = & (\kappa P + \kappa^{\prime} e^{-i\alpha_{\kappa^{\prime}}} R)^{T} (v_{R}f)^{-1} 
(\kappa P + \kappa^{\prime} e^{-i\alpha_{\kappa^{\prime}}} R) 
\nonumber\\
& \simeq & \frac{v_{L}}{\beta} P^{T} f^{-1} P \; ,
\nonumber\\
M_{\nu}^{I} & = & v_{L} e^{i\alpha_{L}} f \; .
\end{eqnarray}
Consequently, the connection between CP violation in the quark 
sector and that in the lepton sector, which is 
made through the phase $\alpha_{\kappa^{\prime}}$, appears only 
at the sub-leading order, $\mathcal{O}
\left( \kappa^{\prime}/\kappa \right)$, thus making this connection 
rather weak. 
We will neglect these sub-leading order terms, and   there is thus only one phase, 
$\alpha_{L}$, that is responsible for all leptonic CP 
violation. 

\begin{figure}[t!]
\begin{tabular}{ccccc}
\includegraphics[scale=0.55]{lepg1} & 
\hspace{0.2in} & 
\includegraphics[scale=0.55]{lepg2} & 
\hspace{0.2in} & 
\includegraphics[scale=0.55]{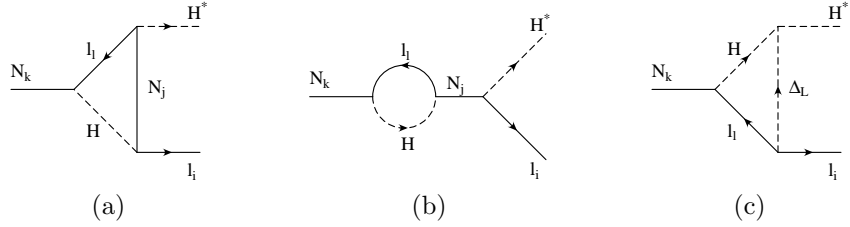} \\
    (a) &  & (b) &  & (c) \\
    \end{tabular}
    \caption{Diagrams in the minimal left-right model that contribute to the lepton number asymmetry through the decay of the RH neutrinos. }
    \label{fig:lepg}
\end{figure}

The three low energy phases $\delta$, $\alpha_{21}$, $\alpha_{31}$, in the MNS matrix are therefore functions of the single fundamental phase, $\alpha_{L}$. Neutrino oscillation probabilities depend on the Dirac phase through the leptonic Jarlskog invariant, which is proportional to $\sin\alpha_{L}$, $J_{CP}^{\ell} \propto \sin\alpha_{L}$. There are two ways to generate lepton number asymmetry. One is  through the decay of the $SU(2)_{L}$ triplet Higgs, $\Delta^{\ast} \rightarrow \ell + \ell$, and the corresponding asymmetry is given by,
\begin{equation}
\epsilon = \frac{\Gamma(\Delta_{L}^{\ast} \rightarrow \ell + \ell) - \Gamma(\Delta_{L} \rightarrow \overline{\ell} + \overline{\ell})}
{\Gamma(\Delta_{L}^{\ast} \rightarrow \ell + \ell) +\Gamma(\Delta_{L} \rightarrow \overline{\ell} + \overline{\ell}) } \; .
\end{equation}
The asymmetry can also be generated through the decay of the lightest RH neutrinos, $N_{1} \rightarrow \ell + H^{\dagger}$, and the asymmetry in this case is,
\begin{equation}
\epsilon = \frac{\Gamma(N_{1} \rightarrow \ell + H^{\dagger}) - \Gamma(N_{1} \rightarrow \overline{\ell} + H)}
{ \Gamma(N_{1} \rightarrow \ell + H^{\dagger}) + \Gamma(N_{1} \rightarrow \overline{\ell} + H)} \; .
\end{equation}
Whether $N_{1}$ decay dominates or $\Delta_{L}$ decay dominates depends upon if $N_{1}$ is heavier or lighter than $\Delta_{L}$. 
As the mass of the triplet Higgs is typically at the scale of the LR breaking scale, it is naturally heavier than the lightest RH neutrino. As a result,  $N_{1}$ decay dominates.  With the particle content of this model, there are three diagrams at one loop that contribute to leptogeiesis, as shown in Fig.~\ref{fig:lepg}. 
The contribution from diagram (a) and (b) mediated by charged lepton and Higgs doublet, which appear also in standard leptogenesis with SM particle content, is given by~\cite{Antusch:2004xy},   
\begin{equation}
\epsilon^{N_{1}}  =  
\frac{3}{16\pi }  \biggl( \frac{M_{R_{1}}  }{v^{2}} \biggr)  
\cdot \frac{
\mbox{Im} \biggl(  \mathcal{M}_{D}  \left( M_{\nu}^{I} \right)^{\ast} 
\mathcal{M}_{D}^{T} 
\biggr)_{11} }{ ( \mathcal{M}_{D} \mathcal{M}_{D}^{\dagger} )_{11} }  \; . 
\end{equation}
Now, there is one additional one-loop diagram, Fig.~\ref{fig:lepg} (c), mediated 
by the $SU(2)_{L}$ triplet Higgs.  It contributes to the 
decay amplitude of the right-handed neutrino into a doublet 
Higgs and a charged lepton, which 
gives an additional contribution to the lepton number 
asymmetry~\cite{Antusch:2004xy},   
 \begin{equation}
 \epsilon^{\Delta_{L}}  =  
\frac{3}{16\pi }  \biggl( \frac{M_{R_{1}}  }{v^{2}} \biggr)  
\cdot \frac{
 \mbox{Im} \biggl(  \mathcal{M}_{D}  \left( M_{\nu}^{II} \right)^{\ast} 
\mathcal{M}_{D}^{T} 
 \biggr)_{11} }{ ( \mathcal{M}_{D} \mathcal{M}_{D}^{\dagger} )_{11} }  \; ,
 \end{equation} 
where $\mathcal{M}_{D}$ is the neutrino Dirac mass term in the basis where 
the RH neutrino Majorana mass term is real and diagonal, 
\begin{equation}
\mathcal{M}_{D} = O_{R} M_{D}, \; \quad 
f^{\mbox{\tiny diag}} = O_{R} f O_{R}^{T} \; .
\end{equation}  
Because there is no phase present in either 
$M_{D} = P \kappa$ or $M_{\nu}^{I}$ or $O_{R}$, 
the quantity $ \mathcal{M}_{D} \left( M_{\nu}^{I} \right)^{\ast} 
\mathcal{M}_{D}^{T}$ is real, leading to a vanishing $\epsilon^{N_{1}}$. 
This statement is 
true for {\it any} chosen unitary transformations $U_{L}$ and $U_{R}$ 
defined in Eq.~(\ref{unit}).  
On the other hand, the contribution, 
$\epsilon^{\Delta_{L}}$, due to the diagram mediated by 
the $SU(2)_{R}$ triplet is proportional to $\sin\alpha_{L}$. 

As all leptonic CP violation in this model come from one single origin, that is, the phase in the VEV of the LH triplet, $\left<\Delta_{L}\right>$, strong correlation between leptogenesis and low energy CP violating processes can thus be established. In particular, both  $J_{CP}^{\ell}$ and $\epsilon$ are proportional to $\sin\alpha_{L}$. 

It has been found recently that, by lowering the left-right symmetry breaking scale with an additional $U(1)$ symmetry, the link between CP violation in the quark sector and that in the lepton sector can also be established~\cite{Chen:2006bv}.

\section{Recent Progress and Concluding Remarks}\label{sec:new}


Leptogenesis provides a very appealing way to generate the observed cosmological baryonic asymmetry. It has gained a significant amount of interests ever since the advent of the evidence of non-zero neutrino masses. In this scenario, the baryonic asymmetry is closely connected to the properties of the neutrinos, and the fact that the required neutrino mass scale for successful leptogenesis is similar to the scale observed in neutrino oscillations makes leptogenesis a very plausible source for the cosmological baryonic asymmetry.  Even though there is so far no direct way to test leptogenesis, the search for leptonic CP violation in neutrino oscillations at very long baseline experiments~\cite{Marciano:2001tz} and to look for lepton number violation in neutrinoless double beta decay will inevitably further the credibility of  leptogenesis as a source of the baryon asymmetry.

The recent developments in the subject of leptogenesis have been focused on the role of flavor. Recall that the total asymmetry given in 
Eq.~\ref{eq:e1} have summed over all three flavor indices, 
\begin{equation}
\epsilon_{1} = \sum_{\alpha = e, \mu, \tau} \epsilon^{\alpha\alpha} \; ,
\end{equation}
where $\epsilon^{\alpha\alpha}$ is the CP asymmetry in the $\alpha$-flavor. Correspondingly, previous solutions to the Boltzmann equations have summed over all the three flavors, $e, \; \mu, \; \tau$, and thus they did not include flavor dependence~\cite{Abada:2006ea},
\begin{eqnarray}
\frac{d(Y_{N_{1}}-Y_{N_{1}}^{eq})}{dz} & = &  - \frac{z}{sH(M_{1})} \biggl( \gamma_{D} 
+ \gamma_{\Delta L = 1} \biggr) \biggl( \frac{Y_{N_{1}}}{Y_{N_{1}}^{eq}}-1\biggr) 
\\
&& - \frac{dY_{N_{1}}^{eq}}{dz} \; , \nonumber
\\
\frac{dY_{L}}{dz} &  = &  \frac{z}{sH(M_{1})} 
\biggl[ \biggl( \frac{Y_{N_{1}}}{Y_{N_{1}}^{eq}} - 1 \biggr) \epsilon_{1} \gamma_{D} \label{eq:btzL}
\\
&&\qquad \qquad +  - \frac{Y_{L}}{Y_{L}^{eq}} \biggl( \gamma_{D} 
 \gamma_{\Delta L = 1} + \gamma_{\Delta L = 2} \biggr) \biggr] \nonumber ,
\end{eqnarray}
where $Y_{N_{1}}$ and $Y_{L}$ are the number density of the lightest right-handed neutrino $N_{1}$ and of the lepton number asymmetry, respectively, 
and $\gamma$'s are the decay rates for the processes specified in the subscripts. It has recently  been pointed out that flavor effects matter if heavy neutrino masses are hierarchical~\cite{Abada:2006ea}. The Yukawa interactions of all three flavors, $e$, $\mu$ and $\tau$, reach equilibrium at different temperatures. These temperatures are determined by the size of the Yukawa couplings, $\lambda$, as
\begin{equation} 
\lambda^{2} M_{Pl} = T_{eq} \; .
\end{equation}
Due to the relative large coupling constant, the $\tau$ Yukawa interactions reach equilibrium at $T \sim 10^{12}$ GeV, while the muon Yukawa interactions reach equilibrium at $T \sim 10^{9}$ GeV. If leptogenesis takes place at $T \sim M_{1} > 10^{12}$ GeV, the Yukawa interactions of all three lepton flavors  are out of equilibrium, and hence the three flavors are indistinguishable. In particular, the washout factor is universal for all three flavors. However, if leptogenesis takes  place at temperature below $10^{12}$ GeV, which is generally the case for hierarchical right-handed neutrino masses,  the three flavors are distinguishable and thus their effects should be included in the Boltzmann equations properly. Instead of a single evolution function for $Y_{L}$ as given in Eq.~\ref{eq:btzL},  one should consider the evolution of the lepton number asymmetry, $Y^{\alpha\alpha}$, which is due to the decay of the lightest right-handed neutrino into charged lepton of flavor $\alpha$ with the corresponding asymmetry given by $\epsilon^{\alpha\alpha}$ and decay rate given by $\gamma_{D}^{\alpha\alpha}$~\cite{Abada:2006ea},
\begin{eqnarray}
\frac{dY^{\alpha\alpha}}{dz} & = & \frac{z}{sH(M_{1})} \biggl[ \biggl( \frac{Y_{N_{1}}}{Y_{N_{1}}^{eq}}-1\biggr) \epsilon^{\alpha\alpha} \biggl( \gamma_{D}^{\alpha\alpha} + \gamma_{\Delta L = 1} \biggr) \\
&& - \frac{Y^{\alpha\alpha}}{Y_{L}^{eq}} \biggl( \gamma_{D}^{\alpha\alpha} + \gamma_{\Delta L = 1} \biggr) \biggr] \; , \nonumber
\end{eqnarray}
Note that in the above equation, there is no summation over the flavor index, $\alpha$. By properly including the flavor effects, the amount of leptogenesis may be enhanced by a factor of $2$ to $3$~\cite{Abada:2006ea}. 

Except for the specific types of models~\cite{Frampton:2002qc,Chen:2004ww} discussed in Sec.~\ref{sec:connect}, the general lack of connection between leptogenesis and low energy CP violation translates into the fact that the observation of the leptonic Dirac or Majorana phases at low energy does not imply non-vanishing leptogenesis. This statement is weakened in a framework when the right-handed neutrino sector is CP invariant and when the flavor effects are important~\cite{Nardi:2006fx}.  
This is elucidate  by introducing the ``orthogonal parametrization'' for neutrino Dirac Yukawa matrix~\cite{Casas:2001sr},
\begin{equation}
h = \frac{1}{v} M^{1/2} R m^{1/2} U^{\dagger} \; ,
\end{equation}  
where $m = \mbox{diag}(m_{1},m_{2},m_{3})$ is the diagonal matrix of the light neutrino masses, $M$ is the diagonal matrix of the right-handed neutrino masses and $U$ is the MNS matrix. The orthogonal matrix $R$ is defined by this equation  as  $R = v M^{-1/2} h U m^{-1/2}$.  
In the basis where the right-handed neutrino mass matrix and the charged lepton mass matrix are diagonal, the neutrino Dirac Yukawa matrix can be written as $h = V_{R}^{\nu\, \dagger} \mbox{diag}(h_{1},h_{2},h_{3})V_{L}^{\nu}$. Therefore, the low energy CP violation in the lepton sector can arise from either the left-handed sector through $V_{L}^{\nu}$, the right-handed sector through $V_{R}^{\nu}$, or from both. 
From $hh^{\dagger} v^{2} 
=  V_{R}^{\nu\, \dagger} \mbox{diag}(h_{1}^{2},h_{2}^{2},h_{3}^{2}) V_{R}^{\nu} v^{2}= M^{1/2} R m R^{\dagger} M^{1/2}$, it can be seen that the phases of $R$ are related to those in the right-handed sector through $V_{R}^{\nu}$. 
The asymmetry $\epsilon_{1}$ given in Eq.~\ref{eq:e1}, which is derived with one-flavor approximation, can be rewritten as follows~\cite{Pascoli:2006ie},
\begin{equation}
\epsilon_{1} = -\frac{3M_{1}}{16\pi v^{2}} \frac{\mbox{Im}\bigl( \sum_{\rho} m_{\rho}^{2} R_{1\rho}^{2}\bigr)}{\sum_{\beta} m_{\beta} |R_{1\beta}|^{2}} \; .
\end{equation}
 Assuming the right-handed sector is CP invariant,  low energy CP phases can then arise entirely from the left-handed sector and thus are irrelevant for $\epsilon_{1}$, which vanishes because the orthogonal matrix $R$ is real.  
If leptogenesis takes place at $T < 10^{12}$ GeV, the flavor effects must be taken into account. In this case the asymmetry in each flavor is given by~\cite{Pascoli:2006ie},
\begin{equation}\label{eq:e3}
\epsilon_{\alpha} = -\frac{3M_{1}}{16\pi v^{2}} \frac{\mbox{Im} \bigl( \sum_{\beta\rho} m_{\beta}^{1/2} m_{\rho}^{3/2} U_{\alpha\beta}^{\ast} 
U_{\alpha \rho} R_{1\beta} R_{1\rho}\bigr)}{\sum_{\beta} m_{\beta} |R_{1\beta}|^{2}} \; .
\end{equation} 
The contribution of each of these individual asymmetries to the total asymmetry is then weighted by the corresponding washout factor.  
Therefore, barring accidental cancellations, the presence of the MNS matrix elements in Eq.~\ref{eq:e3} signifies the need for low energy CP violation in order to have leptogenesis. Hence if leptonic CP violation in neutrino oscillations is observed at future very long baseline experiments~\cite{Marciano:2001tz} and if lepton number violation is established by observing neutrinoless double beta decay, it would even more strongly suggest  than it has been that leptogenesis be the source for the origin of the cosmological baryon asymmetry. 

Finally, a fundamental problem in the current treatment of leptogenesis is the fact that the Boltzmann equations utilized in the present calculations are purely classical treatment. However, the collision terms are zero-temperature S-matrix elements which involve quantum interference. In addition, the time evolution of the system should be treated quantum mechanically. These lead to the need of quantum Boltzmann equations which is based on Closed-Time-Path ({\bf CTP}) formalism~\cite{Mahanthappa:1962ex}. A more detailed discussion on this issue can be found in Ref.~\cite{Riotto:1998bt,DeSimone:2007rw}.


\section*{Acknowledgments}

I would like to thank the organizers, Sally Dawson, K.T. Mahanthappa and Rabi Mohapatra for inviting me to lecture and for organizing such an intellectually stimulating TASI summer school, which has been a very essential experience for graduate students in theoretical high energy physics. I would also like to thank the student participants for their interesting questions and for their enthusiastic participation at the school.

\printindex                         
\end{document}